\documentclass[twocolumn,showkeys,showpacs,preprintnumbers,prd,superscriptaddress,nofootinbib]{revtex4}
\usepackage{graphicx}
\usepackage{epsf}
\usepackage{bm}
\usepackage{amsmath}
\usepackage{amsfonts}
\usepackage{amssymb}
\usepackage{epstopdf}
\usepackage{natbib}
\usepackage{hyperref}
\usepackage{color}
\usepackage{verbatim}
\usepackage{multirow}
\usepackage{bm}
\usepackage{hyperref}
\usepackage{listings}
\usepackage{soul}

\begin{document}


\title{Unveiling the Hubble Constant through Galaxy Cluster Gas Mass Fractions}

\author{Javier E. Gonzalez} \email{javiergonzalezs@academico.ufs.br}
\affiliation{Departamento de F\'{i}sica, Universidade Federal de Sergipe, São Cristóvão, SE 49100-000, Brazil}

\author{Marcelo Ferreira}
\email{fsm.fisica@gmail.com}
\affiliation{Universidade Federal do Rio Grande do Norte, Departamento de F\'{i}sica Te\'{o}rica e Experimental, 59300-000, Natal - RN, Brazil.}

\author{Leonardo R. Cola\c{c}o}
\email{colacolrc@gmail.com}
\affiliation{Instituto Federal da Paraíba (IFPB) - Campus Catolé do Rocha, 58884-000, Catolé do Rocha - PB, Brazil.}

\author{Rodrigo F. L. Holanda}
\email{holandarfl@gmail.com}
\affiliation{Universidade Federal do Rio Grande do Norte, Departamento de F\'{i}sica Te\'{o}rica e Experimental, 59300-000, Natal - RN, Brazil.}

\author{Rafael C. Nunes}
\email{costa.nunes@ufrgs.br}
\affiliation{Instituto de F\'{i}sica, Universidade Federal do Rio Grande do Sul, 91501-970 Porto Alegre RS, Brazil}
\affiliation{Divis\~ao de Astrof\'isica, Instituto Nacional de Pesquisas Espaciais, Avenida dos Astronautas 1758, S\~ao Jos\'e dos Campos, 12227-010, SP, Brazil}

\begin{abstract}
  
In this work, we obtain Hubble constant ($H_0$) estimates by using two galaxy cluster gas mass fraction measurement samples, Type Ia supernovae luminosity distances, and the validity of the cosmic distance duality relation. Notably, the angular diameter distance (ADD) to each galaxy cluster in the samples is determined by combining its gas mass fraction measurement with galaxy clustering observations, more precisely, the $\Omega_b/\Omega_m$ ratio. Such a combination results in a $H_0$ estimate that is independent of a specific cosmological framework. In one of the samples, the gas fraction measurements were calculated in spherical shells at radii near $r_{\rm 2500}$ (44 data points), while in the other (103 data points) the measurements were calculated within $ r_{\rm 500}$. 
We find $H_0=72.7^{+6.3}_{-5.6}$ km/s/Mpc at 68\% CL for the joint analysis of these data sets. We also investigate the impact on the $H_0$ determination by exploring the precision and number of gas mass fraction data by performing a data Monte Carlo simulation. Our simulations show that future measurements could achieve a precision of up to 5\% for $H_0$. 
  

\end{abstract}

\maketitle

\section{Introduction}

In contemporary cosmology, the Hubble constant, denoted as \( H_0 \), holds a pivotal position in our comprehension of the universe's expansion and physical composition. Serving as a fundamental parameter, it delineates the rate at which the universe is currently expanding, thus playing a crucial role in theoretical frameworks aimed at deciphering its age and dynamics. However, recent years have witnessed a surge in interest and discourse concerning the precise value of the Hubble constant. Observations employing various methodologies have produced subtly divergent outcomes, resulting in tensions in the determination of \( H_0 \) \cite{riess2023local,freedman2023progress}.

One of the most significant disparities in cosmological measurements arises between the Planck-CMB estimate \cite{Planck:2018vyg}, based on the standard $\Lambda$CDM model, and the direct local distance ladder measurements by the SH0ES team \cite{Riess:2021jrx,Riess_2022,murakami2023leveraging}, with a significance exceeding 5$\sigma$. This discrepancy is further underscored by various late-time measurements, such as those discussed in \cite{DiValentino:2021izs}, which tend to support a higher value for $H_0$ and are at odds with the Planck-CMB estimate. Furthermore, other late-time measurements also support high values of $H_0$ \cite{DiValentino:2021izs}, while conversely, the lower value of $H_0$ inferred from Planck-CMB data aligns well with constraints derived from Baryon Acoustic Oscillations (BAO) \cite{desicollaboration2024desi,eBOSS:2020yzd}, as well as with results from other CMB experiments \cite{ACT:2020frw,Dutcher:2021vtw,SPT-3G:2022hvq}. Given the persistence of these discrepancies, which cannot be entirely attributed to systematic errors \cite{Riess:2024ohe}, there has been widespread discourse in the literature regarding whether new physics beyond the standard cosmological model may resolve the $H_0$ tension \cite{DiValentino:2021izs,Abdalla:2022yfr,Perivolaropoulos_2022}.

Moreover, it is crucial to measure $H_0$ independently of both CMB data and the local distance ladder method. In this line, one may obtain the Hubble constant via galaxy cluster observations. For instance, in \cite{2012JCAP...02..035H,2014MNRAS.443L..74H} it is considered galaxy cluster angular diameter distance combined with other geometrical measurements to obtain $H_0$ with an accuracy of 5\%. Still in this context, precise measurements of $f_{\text{gas}}$ from X-ray data offer a pathway to constrain cosmological models. When combined with external data, they have yielded robust constraints on cosmological parameters. Concerning $H_0$, a comprehensive analysis was conducted in \cite{2020JCAP...09..053H}, combining galaxy cluster X-ray gas mass fraction and BAO measurements to derive precise constraints on $H_0$ within both the $\Lambda$CDM and $w$CDM models. Additionally, \cite{Mantz_2021} presents related work in this area. Numerous other analyses utilizing galaxy clusters have been documented in the literature (see, for instance,  \cite{2014MNRAS.443L..74H,2012GReGr..44..501H,2007MNRAS.379L...1C,2023EPJC...83..274B,2019A&A...621A..34K,barbosa2024assessing,panchal2024comparison,Darragh_Ford_2023,Wicker_2023}).


Recently, \cite{Colaco:2023gzy} introduced new constraints on $H_0$ by utilizing a combination of the Pantheon Type Ia supernova (SNIa) sample, galaxy cluster angular diameter distances, and the cosmic distance duality relation validity \textbf{  $ (1+z)^2  D_A / D_L = 1 $} \cite{1933PMag...15..761E}. Their analysis involved aggregating statistical and systematic errors in galaxy cluster measurements in quadrature, resulting in $H_0 = 67.22 \pm 6.07$ km/s/Mpc at the $1\sigma$ confidence level (CL). This methodology closely follows the approach previously established in Ref. \cite{Renzi:2020fnx}, ensuring full independence from any specific cosmological model.

Our study aims to constrain the $H_0$ parameter by leveraging a combination of independent datasets. We utilize unanchored luminosity distance estimates obtained from the apparent magnitude Pantheon SNIa sample along with two galaxy cluster angular diameter distance (ADD) samples. It is important to stress that the galaxy cluster ADDs are obtained by combining  $f_{\text{gas}}$ measurements and galaxy clustering observations. These ADD estimates require the $\Omega_b/\Omega_m$ ratio value, so we incorporate two measurements: one from Planck Collaboration \cite{Planck:2018vyg} and another from galaxy clustering \cite{2024arXiv240319236K}. It is crucial to note that the combination $f_{\text{gas}}$ plus galaxy clustering provides galaxy cluster angular diameter distances directly, being, therefore, fully independent of any specific cosmological model. As previously mentioned, our methodology closely follows that developed in \cite{Renzi:2020fnx}, ensuring complete independence from any particular cosmological model, only the cosmic distance duality relation validity is required. The subsequent section introduces the datasets used in our analysis and the methodology. We then proceed to discuss the statistical construction of our observables. Our findings regarding the constraints on $H_0$ are discussed in Section \ref{Constraints}, followed by concluding remarks in Section \ref{final}.

\section{Data and Methodology}
\label{Methodology}

As mentioned previously, the method relies on unanchored luminosity distances, denoted as $[H_0 {D_L}(z)]^{\rm SN}$, and  angular diameter measurements, $[D_A(z)]^{\text{GC}}$,  from galaxy clusters  at the same redshifts. The data considered in this analysis are:

\begin{itemize}

\item Gas mass fraction sample 1 \textbf{(GMF1)}: we utilize a new sample of $f_{\text{gas}}$ measurements spanning the redshift range $0.018 \leq z \leq 1.160$, compiled in \cite{Mantz_2021}. This dataset comprises 44 massive, hot, and morphologically relaxed galaxy clusters observed by the Chandra telescope (see Fig. \ref{fig:fgas}). Notably, the selection of relaxed systems aims to minimize systematic uncertainties and scatter arising from deviations in hydrostatic equilibrium and spherical symmetry. Additionally, the gas mass fraction of these structures was derived within spherical shells at radii near $r_{2500}$ ($0.8-1.2 r_{2500}$), rather than integrated across all radii ($< r_{2500}$). Here, $r_{2500}$ denotes the radius at which the mean enclosed mass density equals $2500$ times the universe's critical density at the cluster's redshift.  

\item Gas mass fraction sample 2 \textbf{(GMF2)}: We analyze a dataset consisting of 103 observations spanning a redshift range from $0.0473 \geq z \geq 1.235$. This dataset comprises 12 clusters with $z < 0.1$ from the X-COP survey \cite{eckert2019non}; a set of 44 clusters within the range $0.1 \geq z \geq 0.3$ \cite{ettori2010mass}; and observations at high redshifts, consisting of 47 clusters obtained by \cite{ghirardini2017evolution} in the range $0.4 \geq z \geq 1.235$ (see Fig. \ref{fig:fgas}). Here, different from the previous sample, it is important to comment that the $f_{\text{gas}}$ measurements of this sample were  calculated within $r_{500}$, which represents the radius of the cluster where the density of the medium exceeds 500 times the critical energy density. This dataset was curated by \cite{corasaniti2021cosmological}.

\item \textbf{Type Ia supernovae}: Distance modulus measurements of Type Ia supernovae are derived from the Pantheon sample \cite{Scolnic:2017caz}. This dataset includes 1048 light curves, with determined apparent magnitudes. The observations span a redshift range from 0.01 to 2.3. In all our analyses, we consider the apparent magnitudes of the Pantheon SNe Ia sample transformed into uncalibrated luminosity distances.
\end{itemize}


The methodology for constraining the Hubble constant relies on the validity of the cosmic distance duality relation (CDDR). This relation is upheld if the photon number remains conserved along null geodesics in a Riemannian space-time between the observer and the source. Over recent years, various astronomical datasets have been scrutinized to test this relation, with no significant statistical evidence supporting a violation of the CDDR \cite{2012A&A...538A.131H,2022ApJ...939..115X,2019ApJ...885...70L,2021EPJC...81..903L}. Consequently, the broad applicability of this relation underscores its fundamental importance in observational cosmology. Any departure from it could signal the presence of new physics or systematic errors in observations \cite{CDDR,Bassett:2003vu,Ellis2007}. Following \cite{Renzi:2020fnx}, one can write 

\begin{equation}
\label{H0}
H_0 = \frac{1}{(1+z)^2} \frac{[H_0 D_L (z)]^{\rm SN}}{[D_A (z)]^{\text{GC}}}.
\end{equation}

In this context, $[H_0 D_L (z)]^{\text{SN}}$ represents the unanchored luminosity distance (see Fig. \ref{fig:H0DL}). Notably, it becomes feasible to derive estimates for $H_0$ by utilizing measurements of unanchored luminosity distance [$H_0 D_L (z)$]$^{\text{SN}}$ alongside angular diameter distance $[D_A (z)]^{\text{GC}}$ at the same redshift $z$. To achieve that, we employ Type Ia supernovae to derive [$H_0 D_L (z)$]$^{\text{SN}}$ and leverage X-ray gas mass fraction data from galaxy clusters to determine $[D_A (z)]^{\text{GC}}$. Further elaboration on this methodology will be provided in the following sections.

\subsection{Angular diameter distance from galaxy cluster gas mass fraction}

In our analyses, we focus on the gas mass fraction observed in galaxy clusters, defined as $f_{\text{gas}} = M_{\text{gas}} / M_{\text{tot}}$, where $M_{\text{gas}}$ represents the mass of intracluster gas and $M_{\text{tot}}$ denotes the total mass, encompassing both baryonic gas and dark matter. It is worth noting that baryonic matter within galaxy clusters is primarily composed of the X-ray-emitting intracluster gas \cite{1986RvMP...58....1S}. The gas mass $M_{\text{gas}} (<R)$ observed within a radius $R$ through X-ray observations can be expressed as \cite{sarazin1986x}:
\begin{eqnarray}
M_{\text{gas}} (<R) &=& \left( \frac{3 \pi \hbar m_e c^2}{2 (1+X) e^6}
\right)^{1/2}  \left( \frac{3 m_e c^2}{2 \pi k_B T_e} \right)^{1/4}
m_H \nonumber\\
& & \mbox{\hspace{-2.5cm}} \times \frac{1}{[\overline{g_B}(T_e)]^{1/2}}
{r_c}^{3/2} \left
[ \frac{I_M (R/r_c, \beta)}{I_L^{1/2} (R/r_c, \beta)} \right] [L_X
(<R)]^{1/2}\;,
\end{eqnarray}
where $m_e$ and $m_H$ are the electron and hydrogen masses, respectively, $\overline{g_B}(T_e)$ is the Gaunt factor, $X$ is the hydrogen mass fraction, $T_e$ is the gas temperature,  $r_c$ stands for the core radius,  $L_X(<R)$ is the total X-ray luminosity,
$$
I_M (y, \beta) \equiv \int_0^y (1+x^2)^{-3 \beta/2} x^2 dx\;, 
$$
and
$$
I_L (y, \beta) \equiv \int_0^y (1+x^2)^{-3 \beta} x^2 dx\;.
$$
 
This equation is derived by adopting a spherical $\beta$-model for the cluster density profile as \cite{cavaliere1978distribution}:

\begin{equation}
\label{modelbeta}
    n_e(\theta) =  n_0    \left(  1 +  \frac{\theta^2}{\theta_c^2} \right)^{-3 \beta/2},
\end{equation}

\noindent where $\beta$ is a dimensionless parameter that relates the kinetic energy of the galaxies within a cluster to the thermal energy of the hot gas that permeates the cluster, $\theta$ represents the radial distance from the center of the cluster,
$\theta_c$ is the core radius and $n_0$ is the central electron density \cite{sarazin1986x}.

By assuming that the intracluster gas is in hydrostatic equilibrium, the total mass within a given radius $R$ can be obtained via \cite{2011ARAA..49..409A}
\begin{equation}
    M_{\text{tot}}(<R) = -\left. \frac{k_B T_e R}{G \mu m_H} \frac{d \ln n_e(r)}{d \ln r} \right|_{r=R}.
\end{equation}

\begin{figure}[htbp]
    \centering
    \includegraphics[scale=0.44]{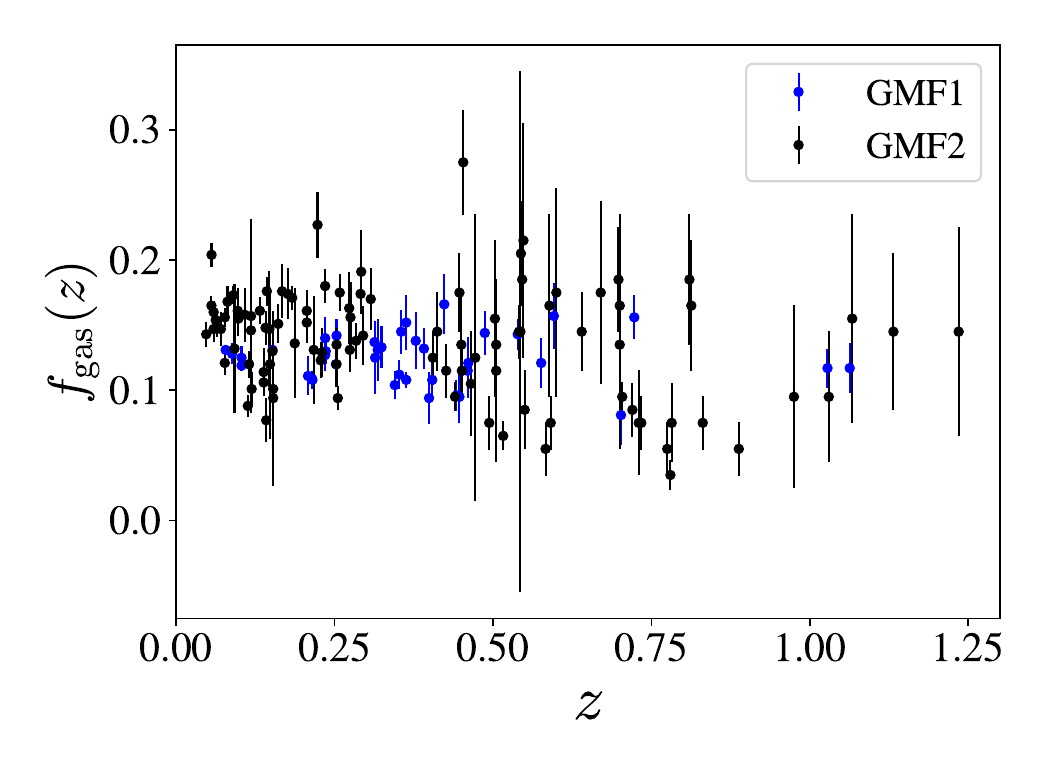}
    \caption{Gas mass fraction data, $f_{\text{gas}}$, for the two samples considered in this work.}
    \label{fig:fgas}
\end{figure}

The expected constancy of $f_{\text{gas}}$ within massive, hot, and relaxed galaxy clusters is commonly utilized to constrain cosmological parameters. In analyses using the galaxy cluster sample from Ref. \cite{Mantz_2021}, cosmological studies are achieved using the following equation (referenced in, for example, \cite{2008MNRAS.383..879A, 2009A&A...501...61E, 2011ARA&A..49..409A, Mantz:2014xba, 2019JCAP...11..032H, Mantz_2021}):

\begin{equation}
\label{fgas1}
f_{\text{gas}}(z) = \gamma_g(z)K(z) A(z) \left[\frac{\Omega_b}{\Omega_m}\right] \left(\frac{D_A^*}{D_A}\right)^{3/2}.
\end{equation}
In this expression, $D_A^*$ denotes the angular diameter distance to the galaxy cluster used in the observations to obtain  $f_{\text{gas}}$ (a flat-$\Lambda$CDM model with Hubble constant $H_0=70$ km s$^{-1}$ Mpc$^{-1}$ and the present-day total matter density parameter $\Omega_m=0.3$),  $K(z)$ stands for the mass calibration and $A(z)$ represents the angular correction factor, which is close to unity for all cosmologies and redshifts of interest, and it can be neglected without significant loss of accuracy \cite{2008MNRAS.383..879A}. This quantity is independent of the Hubble constant. The $\gamma_g(z)$ factor is the gas depletion parameter, which indicates the amount of gas that is thermalized within the cluster potential (\cite{2008MNRAS.383..879A,2009A&A...501...61E,2011ARA&A..49..409A,Mantz:2014xba,Mantz_2021}).  The restriction of selecting hot and relaxed clusters is crucial for reducing uncertainties in predicting the $\gamma_g(z)$ factor from hydrodynamic simulations and minimizing intrinsic scatter, thereby leading to tighter cosmological constraints (for further details, see \cite{Mantz_2021}).
It is worth to comment that cosmological analyses with gas mass fraction measurements are model-independent due to the ratio in the parenthesis of Eq.~(\ref{fgas1}), which takes into account the expected variation in the gas mass fraction measurement when the underlying cosmology is varied.

In analyses using the GMF2, where we followed the modeling of the $f_{\text{gas}}$ given by \cite{allen2008improved},

\begin{equation}
    f_{\text{gas}}(z) = K\gamma_b(z) \left( \frac{\Omega_b}{\Omega_m} \right) \left[ \frac{D_A^{*}(z)}{D_A(z)}\right]^{3/2} - f_{*},
    \label{EqGMF2}
\end{equation}
where $f_{*}$ is the stellar fraction. We assume a Gaussian prior on the stellar fraction, such as $f_* = 0.015 \pm 0.005$.  The value of $\gamma_b(z)$  to GMF2 sample corresponds to the baryonic depletion factor and not the gas depletion. Therefore, it is necessary in this case to subtract the mass present in the stars in Eq.\eqref{fgas1}. This value is consistent with estimates from a sample of clusters with masses in the same range as those of the gas mass fraction dataset \cite{eckert2019non}. The GMF1 and GMF2 samples are shown in Fig. \ref{fig:fgas}.

From  Eqs.\eqref{fgas1} and \eqref{EqGMF2} it is straightforward that
\begin{equation}
\label{DA}
D_A(z) = \left(\frac{\gamma_g(z)K(z) A(z) \Omega_b}{f_{\text{gas}}(z)\Omega_m}\right)^{2/3} D_A^*,
\end{equation}
and
\begin{equation}
\label{DA2}
D_A(z) = \left(\frac{\gamma_b(z)K(z)}{(f_{\text{gas}}(z)+f_*)} \frac{\Omega_b}{\Omega_m}\right)^{2/3} D_A^*.
\end{equation}

\begin{figure}
\includegraphics[scale=0.44]{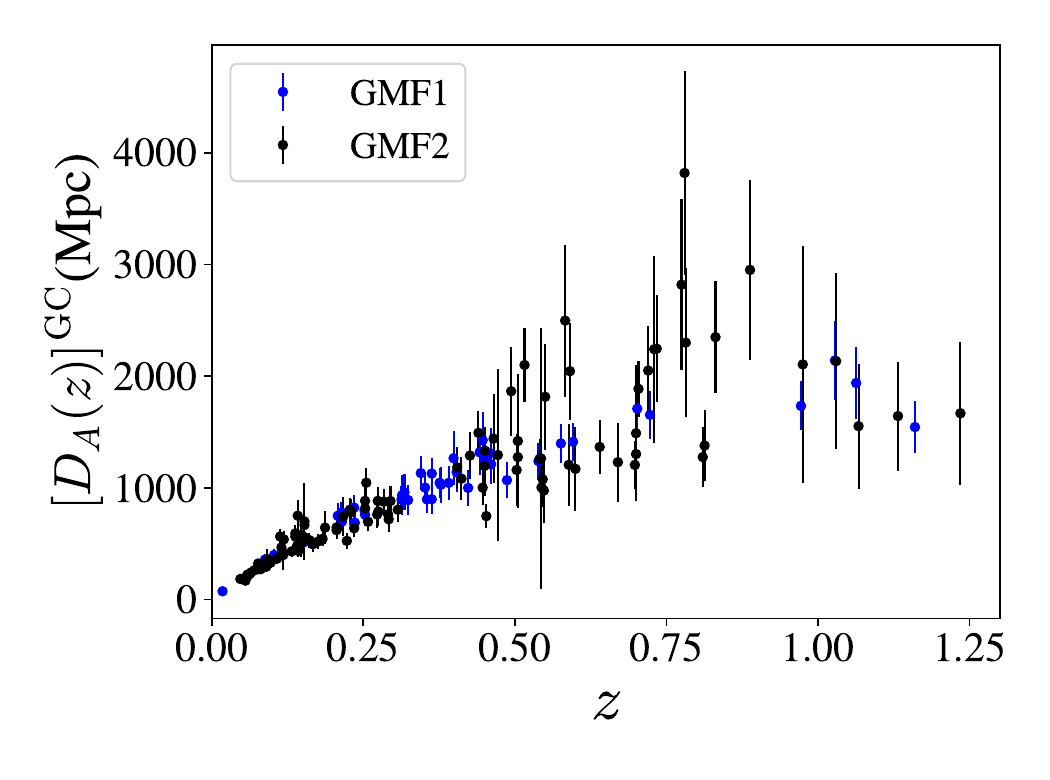}
\caption{Galaxy cluster angular diameter estimates from gas mass fraction measurements (GMF1 and GMF2) by using $\Omega_b/\Omega_M$ obtained via galaxy clustering observations \cite{2024arXiv240319236K}.}
\label{fig:DA}
\end{figure}

The determination of the angular diameter distance relies on the $\Omega_b/\Omega_m$ ratio. To derive the $D_A(z)$ value for each galaxy cluster, we employ two approaches. Firstly, we use the most recent determination of this ratio obtained from galaxy clustering as presented in \cite{2024arXiv240319236K}. Additionally, we perform our analyses by considering this ratio from the Planck-CMB data (TT+TE+EE+lowE+lensing) \cite{Planck:2018vyg}. The combination of $f_{\text{gas}}$ and galaxy clustering provides a direct angular diameter distance for each galaxy cluster.

 For the GMF1 sample, we consider the parameterizations $\gamma_g(z)=\gamma_0(1+\gamma_1z)\left( \frac{M_{2500}}{3 \times 10^{14} M_{\odot}} \right)^\alpha$ (with $\alpha=0.025 \pm 0.033$) and $K(z)=K_0(1+K_1z)$, with uniform priors on $\gamma_1$ and $K_1$ ranging between $-0.05$ and $0.05$ \cite{Mantz:2014xba, Mantz_2021}. Still in agreement with \cite{Mantz:2014xba, Mantz_2021}, we apply the following Gaussian priors: $\gamma_0 = 0.79 \pm 0.07$ \cite{10.1093/mnras/stt265} and $K_0 = 0.93 \pm 0.11$ \cite{10.1093/mnras/stt2129}.  The $\gamma_0$ here is from hydrodynamic simulations and corresponds directly to gas mass fraction (see Table III from \cite{10.1093/mnras/stt265}). The $K_0$ value was obtained by using 13 clusters of the sample by incorporating weak gravitational lensing measurements from the Weighting the Giants project \cite{2014MNRAS.439...48A}. On the other hand, for the GMF2 sample, we use the values $\gamma_b(z) = 0.931(1 + 0.017z + 0.003z^2)$ (with standard deviation $\sigma=0.04$) in full agreement with the results from the FABLE simulations \cite{10.1093/mnras/staa2235} and $K = 0.84 \pm 0.04$ given by the analysis of a sample of clusters in Ref. \cite{10.1093/mnras/staa2303} from the Canadian Cluster Comparison Project \cite{10.1093/mnras/stv275}.  By considering the Eq.\eqref{DA2} from the Ref.\cite{Angelinelli:2022njb}, we have multiplied this factor by $w^{\beta}x^{y+\delta * w}$, where  $w = M_{500} / 5\cdot10^{14} \ h^{-1}M_{\odot}$, $x = r /R_{500, \mathrm c}$, while $\beta$, $\gamma$ and $\delta$ are the free parameters are $(\beta,\gamma,\delta)= (0.12, 0.23, -0.22)$. This takes into account a possible radial and mass dependency of $\gamma_b(z)$. Naturally, the values adopted are different for each GMF sample due to the x-ray gas mass fraction measurements having been obtained by considering different galaxy cluster radii. The angular diameter estimates from these two samples are presented in Fig. \ref{fig:DA}, where we consider the previously mentioned mean values and errors of the astrophysical parameters and the $\Omega_b/\Omega_M$ obtained via galaxy clustering observations \cite{2024arXiv240319236K}.

\subsection{The unanchored luminosity distance from Pantheon SN Ia compilation}

We  require  the unanchored luminosity distance $[H_0 D_L (z)]$, which can be obtained from the apparent magnitude of SNe Ia through the following relation:

\begin{equation}
\label{m_B}
m_B = 5 \log_{10} [H_0 D_L (z)] - 5a_B,
\end{equation}
where $m_B$ is the apparent peak magnitudes of the Pantheon catalog, and we consider the intercept estimate of the Hubble diagram, $a_B = 0.71273 \pm 0.00176$  \cite{Riess:2016jrr}. Such $a_B$ value is independent of any absolute scale of luminosity or distance, and its combination with the apparent magnitude provides distance measurements independent of the selection of the light-curve fitter, fiducial source or filter. This quantity has a weak dependence on  cosmological models, with the estimate based on the cosmographic parameters $q_0$=-0.55 and $j_0$=1 \cite{Riess:2016jrr}.

To obtain a sample of uncalibrated luminosity distances, we transform the Pantheon apparent magnitude measurements with the following relation \cite{Colaco:2023gzy}:

\begin{equation}
\label{unanchored_luminosity}
[H_0 D_{L}]^{\text{SN}}(z) = 10^{(m_b + 5a_B)/5}\equiv 10^{m_b'/5}.
\end{equation}

\noindent In order to take into account the complete systematics and statistical uncertainties of the SN data in our analysis, we add the error of the intercept of the Hubble diagram to  the non-diagonal covariance matrix of the apparent magnitude,

\begin{equation}
    \textbf{Cov} (\bm m'_b)=\textbf{Cov}(\bm m_b)+(5\sigma_{a{_B}})^2\bm I,
\end{equation}
being $\bm I$ the unity matrix. Thus, we apply the matrix transformation relation of the $m'_B$ into the unanchored luminosity distance,
 \begin{equation*}
     \textbf{Cov}([\bm H_0 \bm D_{L}]^{\text{SN}}) =
 \end{equation*}
\begin{equation}
\label{cov_H0DL}
\left(\frac{\partial  [\bm H_0 \bm D_{L}]^{\text{SN}} }{\partial \bm m'_b}\right)\textbf{Cov}(\bm m'_b) \left(\frac{\partial  [\bm H_0 \bm D_{L}]^{\text{SN}} }{\partial \bm m'_b}\right)^T,
\end{equation}
where $\frac{\partial  [\bm H_0 \bm D_{L}]^{\text{SN}} }{\partial \bm m'_b}$ represents the partial derivative matrix of the unanchored luminosity distance vector $[\bm H_0 \bm D_{L}]^{\text{SN}} $ concerning the vector $\bm m'_b$. Because the luminosity distance of a SN Ia depends only on its own apparent magnitude, this matrix is diagonal.

\begin{figure}
\includegraphics[scale=0.5]{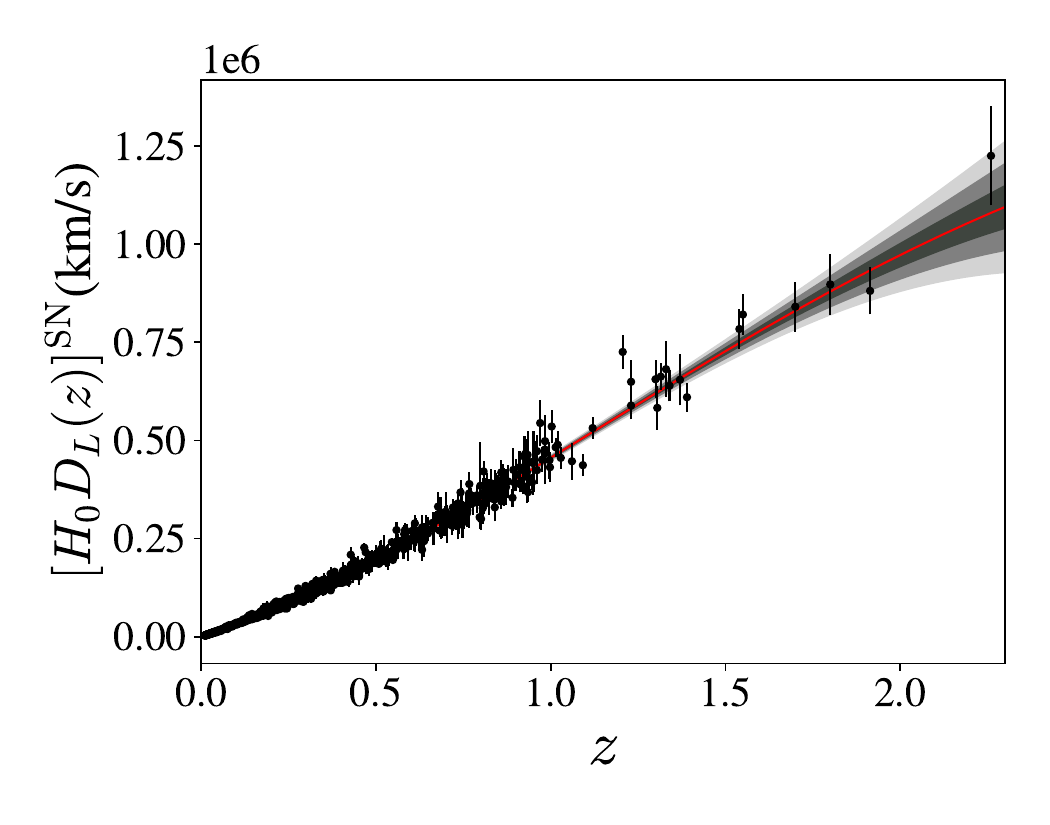}
\caption{The GP reconstruction of $[H_0 D_L (z)]^{\text{SN}}$ from the SN Ia Pantheon data compilation.}
    \label{fig:H0DL}
\end{figure}

As it was explained, this method to estimate the Hubble constant requires measurements of $[ H_0 D_{L}]^{\text{SN}} $ and $[ D_{A}]^{\text{GC}} $ at the same redshift.  To obtain the unanchored luminosity distance at the redshift of gas mass fraction data, we perform a  Gaussian Process (GP) regression of the constructed $[ H_0 D_{L}]^{\text{SN}} $ sample considering the covariance matrix in Eq. (\ref{cov_H0DL}). The GP method assumes a prior mean function to describe the behavior of the data and a covariance function or kernel to quantify the correlation of the observable in different domain points. Due to its continuity and differentiability, we adopt the usual square exponential kernel given by:

\begin{equation}
k(z,z')=\sigma^2 \exp\left(-\frac{(z-z')^2}{2l^2}\right),
\end{equation}
where $\sigma$ and $l$ are the hyperparameters related to the variation of the estimated function and its smoothing scale, respectively. We consider the zero function as a prior mean function to reduce the bias that a cosmological model can generate. The final reconstruction is obtained from the conditional probability considering the prior mean function given the data and depends explicitly on the values of these hyperparameters. We consider the values of $\sigma$ and $l$ that maximize the marginal likelihood, or equivalently, its logarithm, 

\begin{equation*}
\ln {\mathcal{L}}=-\frac{1}{2}([\bm {H_0 D_{L}}]^{\text{SN}})^T [\bm K(\bm z,\bm z)+\bm C]^{-1}[\bm {H_0 D_{L}}]^{\text{SN}}
\end{equation*}

\begin{equation}
\label{log_likelihood}
  -\frac{1}{2}\ln |\bm K(\bm z,\bm z)+\bm C|-\frac{n}{2}\ln 2\pi,
\end{equation}

\noindent where $\bm z$ and $[\bm H_0 \bm D_L (z)]^{\text{SN}}$ are the vectors of the independent and dependent data variables, respectively,  $\bm C$ is the covariance matrix of the data given by Eq. (\ref{cov_H0DL}) and $n$ the number of data points.  The mean value of the reconstructed function , $\overline{ [\bm {H_0 D_{L}}]}^*$ ,  given the hyperparameters obtained by maximising Eq. (\ref{log_likelihood}) is
\begin{equation}
    \overline{ [\bm {H_0 D_{L}}]} ^*=\bm K(\bm z^*, \bm z) [\bm K(\bm z,\bm z)+\bm C]^{-1} [\bm {H_0 D_{L}}]^{\text{SN}}.
\end{equation}
In this case, $\bm z^*$ is the redshift vector of the $[ D_{A}]^{\text{GC}} $ data. These $\overline{ [\bm {H_0 D_{L}}]}^*$ estimates are correlated and their covariance matrix is given by:

\begin{equation*}
   \bm C_{H_0D_L} \equiv \textbf {Cov}([\bm {H_0 D_{L}}] ^*)=
\end{equation*}
\begin{equation}
    \bm K(\bm z^*, \bm z) [\bm K(\bm z,\bm z)+\bm C]^{-1}\bm K(\bm z, \bm z^*).
\end{equation}
We use the GaPP\footnote{https://github.com/carlosandrepaes/GaPP}  code to estimate the $\sigma$, $l$ and the  $\overline{ [\bm {H_0 D_{L}}]}^*$ values \cite{Seikel_2012}. In Fig. \ref{fig:H0DL}, we present the GP reconstruction of the unanchored luminosity distance. In what follows, we present the $H_0$ constraints.

\section{Results}
\label{Constraints}

We conduct a joint Bayesian statistical analysis of the two aforementioned gas mass fraction data samples and the Markov Chain Monte Carlo (MCMC) method to constrain $H_0$ with all data available. We employ the emcee sampler \cite{2013PASP125306F} to generate the chains and use the GetDist Python package \cite{Lewis:2019xzd} for chain processing. In order to avoid a double counting in the statistical analysis, we exclude  the common clusters from the GMF2 sample: Abell2029, Abell1835, Abell2204, J1415.1+3612, RXCJ2129.6+0005, RXJ1347.5-1145 and SPT-CLJ2043-5035.

By defining the vector $\bm \Delta \equiv H_0(1+\bm z^*)^2[\bm D_A(\bm z^*)]^{\text{GC}}-[\bm H_0 \bm D_L (\bm z^*)]^{\text{SN}}$, the $\chi^2$ function can be written as 
\begin{equation}
    \chi^2=\bm \Delta^T(\bm C_{ D_A}+\bm C_{H_0D_L})^{-1}\bm \Delta,
\end{equation}
where $\bm C_{D_A}$ is the diagonal matrix of the ADD errors which elements are

\begin{equation}
    (\sigma_{ D_A,i})^2=\left(\frac{2}{3}\frac{H_0(1+z_i)^2[D_A(z_i)]^{\text{GC}}\sigma_{f_{\text{gas}},i}}{f_{{\text{gas}}, i}}\right)^2.
\end{equation}
Finally, we express the joint likelihood function of our analysis as

\begin{equation}
\label{likelihood}
    \mathcal{L}(\mathrm{Data}|{\bm \Theta}) = \frac{1}{(2\pi)^{n/2}|\bm C_{ D_A}+\bm C_{H_0D_L}|^{1/2}} \exp\left({ -\frac{1}{2}\chi^2}\right),
\end{equation}
where ${\bm \Theta}=\{H_0,\omega_b/\omega_m,K_0^{r_{2500}},K_1^{r_{2500}},\gamma_0^{r_{2500}},\gamma_1^{r_{2500}},\alpha,$ $K_0^{r_{500}},\gamma_0^{r_{500}} \}$ is the vector of the free parameters used in the analysis. It is worth to mention that we treat the gas mass calibration parameter, $K$, and the depletion factor parameter, $\gamma$, independently because each one corresponds to  models with different spherical radii.  

It is worth mentioning that in Eq. (\ref{likelihood}), the normalization term in the likelihood relating to the errors of the cosmological observables is not negligible due to the dependence of $\sigma_{\rm DA,i}$  on the set of parameters. Finally, the posterior distribution to be sampled in the MCMC analysis is constructed as
\begin{equation}
    P({\bm \Theta}|\mathrm{Data})  \propto  \mathcal{L} (\mathrm{Data}|{\bm \Theta}) \times \prod_i \pi({\Theta_i}).
\end{equation}
where $\pi({\Theta_i})$ corresponds to the prior probability function considered for each parameter in Table \ref{priors}.

As our primary focus is on constraining the Hubble constant, we have marginalized over all cosmological and astrophysical parameters. Notably, the expressions of $D_A$ in Eqs. (\ref{DA}) and (\ref{DA2}) crucially depend on the ratio $\Omega_b/\Omega_m=\omega_b/\omega_m$. To account for that, we incorporate a Gaussian prior for this ratio derived from galaxy clustering constraints \cite{2024arXiv240319236K}, alongside Gaussian priors for individual parameters provided by the Planck Collaboration \cite{Planck:2018vyg} (see Table \ref{priors}).

\begin{table}[]
    \centering
    \begin{tabular}{c c c}
 
         Parameter & Prior& Observable \\
         \hline 
            \hline 
         \rule{0pt}{3ex} 
                   
          $H_0$(km/s/Mpc)& $\mathcal{U}[20;120]$& - \\
          $\omega_{b,0}/\omega_{m,0}$ & $\mathcal{N}[0.1564, 0.001596]$ &CMB\\
         $\omega_{b,0}/\omega_{m,0}$ & $\mathcal{N}[0.173;0.027]$&Galaxy Clustering\\
            \hline 
    \end{tabular}
    \caption{Priors used in the statistical analysis. Gaussian priors are noted with $\mathcal{N}$[mean; standard deviation], while flat priors are noted as $\mathcal{U}$[min; max]. }
    \label{priors}
\end{table}

\begin{figure}[t]
\includegraphics[scale=0.75]{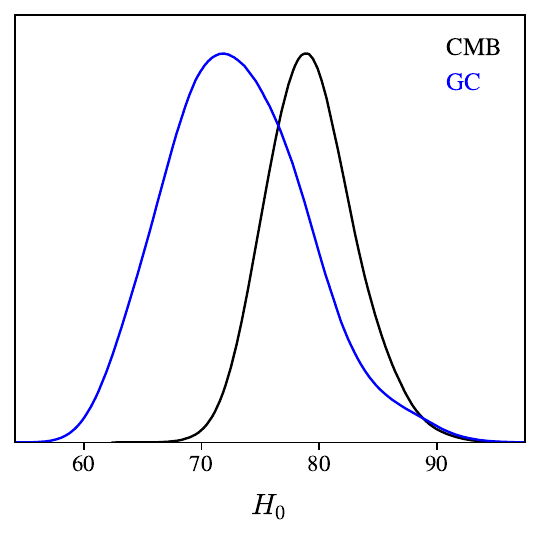}
    \caption{ Posterior distributions of $H_0$ (km/s/Mpc) obtained by employing $\Omega_b$ and $\Omega_m$ priors from Planck 2018 \cite{Planck:2018vyg}, and utilizing the $\Omega_b / \Omega_m$ prior from galaxy clustering \cite{2024arXiv240319236K}.}
    \label{fig:H0}
\end{figure}

The results of the joint analysis of the GMF1 and GMF2 are as follows:

\begin{itemize}

\item  Utilizing the galaxy clustering prior yields $H_0=72.7^{+6.3}_{-5.6}$ km/s/Mpc.

\item  Utilizing the Planck prior, $H_0=79.1^{+4.3}_{-3.6}$ km/s/Mpc.

\end{itemize}

The posteriors are illustrated in Fig. \ref{fig:H0}. These results underscore the significant influence of a precise determination of the ratio $\omega_b/\omega_m$ on the estimation of $H_0$. It is important to note that our primary findings stem from model-independent analyses utilizing galaxy cluster data. Conversely, the determination of the $\omega_b/\omega_m$ ratio from CMB experiments is contingent upon the $\Lambda$CDM model.

 The authors of the Ref.\cite{Mantz_2021} considered  the six clusters with $z< 0.16$ of the GMF1 sample,  $K_0 = 0.93 \pm 0.11$, a Gaussian prior on $\Omega_b/\Omega_m$ from the Planck results (considering the flat $\Lambda$CDM model) and marginalized over a non-flat cosmological model with $\omega$ free  (dark energy equation parameter). It was estimated $H_0=72.22 \pm 6.7$ km/s/Mpc. As one may see, their estimate is in full agreement with ours.  However, no specific cosmological framework was considered in our estimates, we  marginalized on a possible evolution  of the astrophysical parameters or mass dependence ($\gamma(z)$  $K(z)$ and $\alpha$) and considering all galaxy clusters in a joint statistical analysis of the GMF1 and GMF2 samples.

\subsection{Estimating $H_0$ using simulated data}

\begin{figure*}[t]
\includegraphics[scale=0.45]{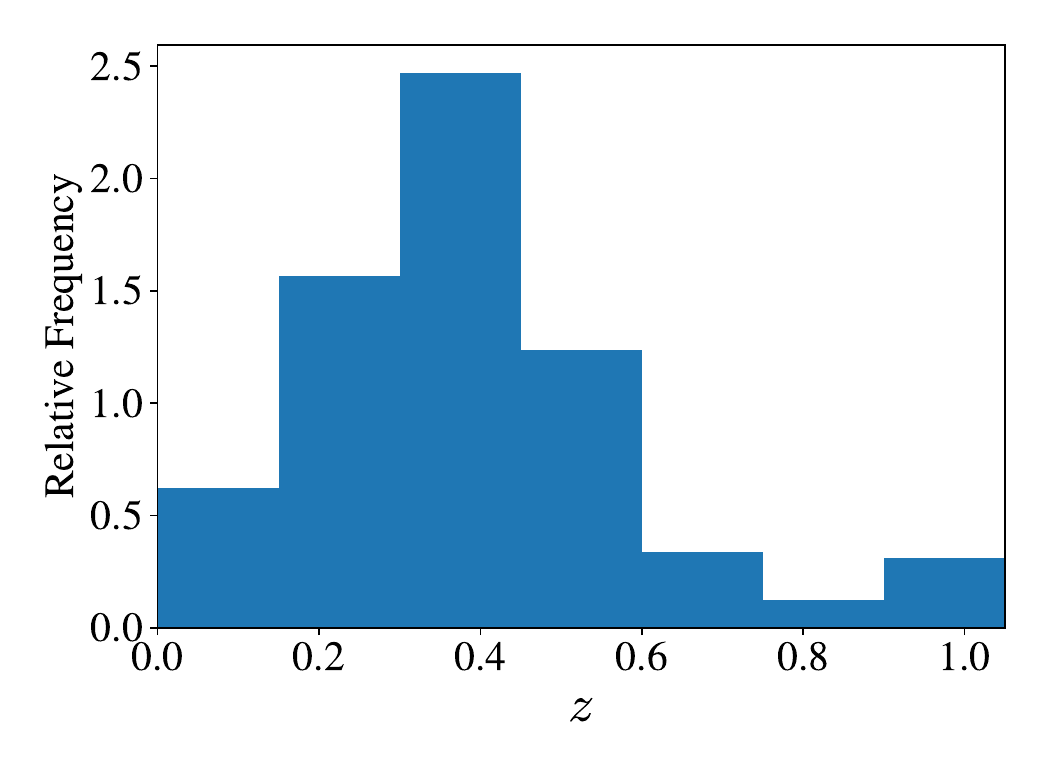} \,\,\,
\includegraphics[scale=0.45]{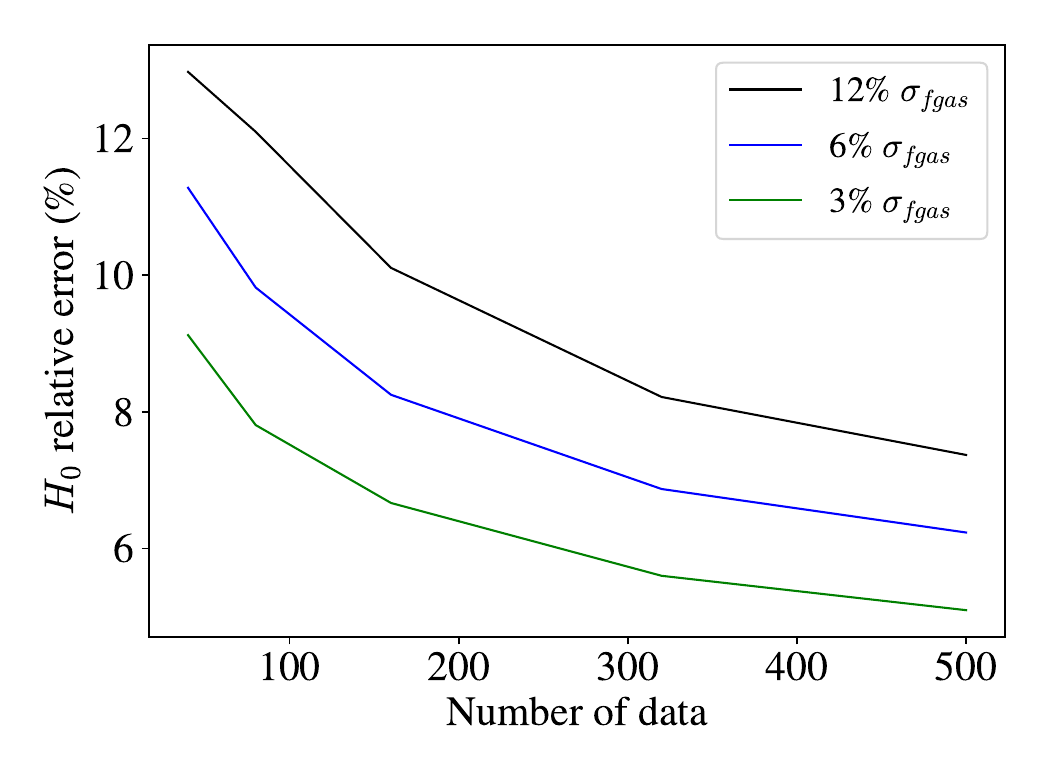}
\caption{\textbf{Left Panel:} Redshift distribution of the GMF1-like simulated data.
\textbf{Right Panel:} Precision in $H_0$ as a function of the number of $f_{\text{gas}}$ observations. Each line represents a specific value of the average relative error of the simulated data.}
    \label{fig:zdist}
\end{figure*}


In order to estimate the impact on the Hubble constant precision of the increasing number of gas mass fraction data and the improvements of its precision, we perform a Monte Carlo simulation of GMF1-like data with different relative errors and numbers of data points and apply the statistical method described above using current SN Ia data and the $\Omega_b/\Omega_m$-GC prior. To simulate the GMF1-like data, we consider the following features of the real samples:
\begin{itemize}
    \item The set of realizations of simulated data has the same redshift distribution as the real data, which is shown in Fig. \ref{fig:zdist}.
    \item We fit the redshift evolution of relative errors of the real sample taking into account the mean value and dispersion of this fit.
    \item  As the gas mass fraction in the galaxy cluster is expected to be constant, in addition to the relative error, we also consider the dispersion of the actual data around its mean value across different redshifts. 
\end{itemize}

Taking into account the characteristics of the real data, we perform 100 simulations with 40, 80, 160, 320, and 500 $f_{\text{gas}}$ data points. Given that the average relative error in the real data is approximately 12\%, we investigate the impact on $H_0$ estimates of improved precision in $f_{\text{gas}}$ data by considering simulated data sets with average relative errors of 12\%, 6\%, and 3\%, while maintaining  their fitted
redshift evolution and astrophysical and cosmological parameters unaltered. The results of $H_0$ simmetrized errors \cite{Schmelling:2000td}  performing this analysis are presented in the right panel of Fig. \ref{fig:zdist}. Assuming that future measurements maintain the same average relative error as current samples, we can project that, with an optimistic scenario of 500 $f_{\text{gas}}$ measurements, an accuracy of 7.3\% in $H_0$ could be achieved. Furthermore, if future $f_\text{gas}$ data precision improves by factors of 2 and 4 over current levels, $H_0$ could be measured with precision of 6.2\% and 5.0\%, respectively. It is worth mentioning that in this analysis, the precision of individual $H_0$ estimates obtained from each $f_{\text{gas}}$ data point is limited by the precision of the astrophysical and cosmological parameters. However, by  combining  all $H_0$ information statistically, the final estimate may achieve higher precisions. A similar conclusion is found for precision in terms of GMF2-like samples. Therefore, for consistency in presentation, we have retained this analysis for our main sample, the GMF1-like data samples. Importantly, in an optimistic observational scenario, we may achieve a competitive $H_0$ accuracy in the near future.

 





\section{Final Remarks}
\label{final}

In this work, we have presented new estimates on the Hubble constant without reliance on any specific cosmological model, utilizing observations of galaxy clusters and SNe Ia. Our approach hinges solely on the validity of the cosmic distance duality relation.

For galaxy clusters, we utilized two distinct gas mass fraction measurement samples: one comprising 44 measurements (GMF1) and the other consisting of 103 measurements (GMF2), obtained from literature sources. Notably, these measurements are derived by considering different radii for these structures: $r_{2500}$ and $r_{500}$, respectively. However, to implement our methodology, we require the angular diameter distance (ADD) for each galaxy cluster. This quantity is derived by combining gas mass fraction measurements with galaxy clustering observations, specifically the $\Omega_b/\Omega_m$ ratio. This combination directly furnishes the ADD for each galaxy cluster, enabling Hubble constant estimates independent of any specific cosmological model.

Our  analysis yields the following Hubble constant estimate: $H_0=72.7^{+6.3}_{-5.6}$ km/s/Mpc for the compilation of the samples GMF1 and GMF2 at 68\% CL (see Fig. \ref{fig:H0}). Due to their large uncertainties, these constraints are consistent with the measurements provided by the SH0ES team and the Planck satellite, indicating potential agreement despite the inherent errors in galaxy cluster observations. It is widely recognized that a deeper understanding of intra-cluster gas and dark matter properties is essential to enhance the utility of galaxy clusters as cosmological probes.    In general, the mass of a galaxy cluster obtained using strong gravitational lensing tends to be greater than the mass derived from the hypothesis of hydrostatic equilibrium. In other words, the hydrostatic equilibrium hypothesis adopted in these studies may be underestimating the gas mass fraction measurements and, consequently, overestimating the Hubble constant estimates. However, the galaxy clusters considered in analyses are limited to the most dynamically relaxed, massive clusters known. In turn, this restriction is critical for reducing systematic from hidrostatical equilibrium and spherical symmetry.

Finally, the impact on the $H_0$ determination via data Monte Carlo simulation was explored.  Our simulations showed that future gas mass fraction measurements could achieve a precision of up to 5\% for $H_0$ (see Fig. 5). Therefore, as statistical and systematic uncertainties in galaxy cluster observations diminish, the methodology employed here, leveraging gas mass fraction measurements, holds promise for improving constraints on the Hubble constant in a minimal model-independent manner.

 
          

\begin{acknowledgments}
\noindent The authors thank the referee and Susana Landau for valuable comments about this work. RFLH thanks the financial support from the Conselho Nacional de Desenvolvimento Cient\'{i}fico e Tecnologico (CNPq, National Council for Scientific and Technological Development) under the project No. 309132/2020-7. RCN thanks the financial support from the CNPq under project No. 304306/2022-3, and the Fundação de Amparo à Pesquisa do Estado do RS (FAPERGS, Research Support Foundation of the State of RS) for partial financial support under project No. 23/2551-0000848-3. LRC also thanks to CNPQ support under the project No. 169625/2023-0. ..
\end{acknowledgments}

\bibliography{PRD}

\begin{thebibliography}{65}%
\makeatletter
\providecommand \@ifxundefined [1]{%
 \@ifx{#1\undefined}
}%
\providecommand \@ifnum [1]{%
 \ifnum #1\expandafter \@firstoftwo
 \else \expandafter \@secondoftwo
 \fi
}%
\providecommand \@ifx [1]{%
 \ifx #1\expandafter \@firstoftwo
 \else \expandafter \@secondoftwo
 \fi
}%
\providecommand \natexlab [1]{#1}%
\providecommand \enquote  [1]{``#1''}%
\providecommand \bibnamefont  [1]{#1}%
\providecommand \bibfnamefont [1]{#1}%
\providecommand \citenamefont [1]{#1}%
\providecommand \href@noop [0]{\@secondoftwo}%
\providecommand \href [0]{\begingroup \@sanitize@url \@href}%
\providecommand \@href[1]{\@@startlink{#1}\@@href}%
\providecommand \@@href[1]{\endgroup#1\@@endlink}%
\providecommand \@sanitize@url [0]{\catcode `\\12\catcode `\$12\catcode `\&12\catcode `\#12\catcode `\^12\catcode `\_12\catcode `\%12\relax}%
\providecommand \@@startlink[1]{}%
\providecommand \@@endlink[0]{}%
\providecommand \url  [0]{\begingroup\@sanitize@url \@url }%
\providecommand \@url [1]{\endgroup\@href {#1}{\urlprefix }}%
\providecommand \urlprefix  [0]{URL }%
\providecommand \Eprint [0]{\href }%
\providecommand \doibase [0]{http://dx.doi.org/}%
\providecommand \selectlanguage [0]{\@gobble}%
\providecommand \bibinfo  [0]{\@secondoftwo}%
\providecommand \bibfield  [0]{\@secondoftwo}%
\providecommand \translation [1]{[#1]}%
\providecommand \BibitemOpen [0]{}%
\providecommand \bibitemStop [0]{}%
\providecommand \bibitemNoStop [0]{.\EOS\space}%
\providecommand \EOS [0]{\spacefactor3000\relax}%
\providecommand \BibitemShut  [1]{\csname bibitem#1\endcsname}%
\let\auto@bib@innerbib\@empty
\bibitem [{\citenamefont {Riess}\ and\ \citenamefont {Breuval}(2023)}]{riess2023local}%
  \BibitemOpen
  \bibfield  {author} {\bibinfo {author} {\bibfnamefont {A.~G.}\ \bibnamefont {Riess}}\ and\ \bibinfo {author} {\bibfnamefont {L.}~\bibnamefont {Breuval}}\ }(\bibinfo {year} {2023})\ \Eprint {http://arxiv.org/abs/2308.10954} {arXiv:2308.10954 [astro-ph.CO]} \BibitemShut {NoStop}%
\bibitem [{\citenamefont {Freedman}\ and\ \citenamefont {Madore}(2023)}]{freedman2023progress}%
  \BibitemOpen
  \bibfield  {author} {\bibinfo {author} {\bibfnamefont {W.~L.}\ \bibnamefont {Freedman}}\ and\ \bibinfo {author} {\bibfnamefont {B.~F.}\ \bibnamefont {Madore}},\ }\href@noop {} {\enquote {\bibinfo {title} {Progress in direct measurements of the hubble constant},}\ } (\bibinfo {year} {2023}),\ \Eprint {http://arxiv.org/abs/2309.05618} {arXiv:2309.05618 [astro-ph.CO]} \BibitemShut {NoStop}%
\bibitem [{\citenamefont {Aghanim}\ \emph {et~al.}(2020)\citenamefont {Aghanim} \emph {et~al.}}]{Planck:2018vyg}%
  \BibitemOpen
  \bibfield  {author} {\bibinfo {author} {\bibfnamefont {N.}~\bibnamefont {Aghanim}} \emph {et~al.} (\bibinfo {collaboration} {Planck}),\ }\href {\doibase 10.1051/0004-6361/201833910} {\bibfield  {journal} {\bibinfo  {journal} {Astron. Astrophys.}\ }\textbf {\bibinfo {volume} {641}},\ \bibinfo {pages} {A6} (\bibinfo {year} {2020})},\ \Eprint {http://arxiv.org/abs/1807.06209} {arXiv:1807.06209 [astro-ph.CO]} \BibitemShut {NoStop}%
\bibitem [{\citenamefont {Riess}\ \emph {et~al.}(2022{\natexlab{a}})\citenamefont {Riess} \emph {et~al.}}]{Riess:2021jrx}%
  \BibitemOpen
  \bibfield  {author} {\bibinfo {author} {\bibfnamefont {A.~G.}\ \bibnamefont {Riess}} \emph {et~al.},\ }\href {\doibase 10.3847/2041-8213/ac5c5b} {\bibfield  {journal} {\bibinfo  {journal} {Astrophys. J. Lett.}\ }\textbf {\bibinfo {volume} {934}},\ \bibinfo {pages} {L7} (\bibinfo {year} {2022}{\natexlab{a}})},\ \Eprint {http://arxiv.org/abs/2112.04510} {arXiv:2112.04510 [astro-ph.CO]} \BibitemShut {NoStop}%
\bibitem [{\citenamefont {Riess}\ \emph {et~al.}(2022{\natexlab{b}})\citenamefont {Riess}, \citenamefont {Breuval}, \citenamefont {Yuan}, \citenamefont {Casertano}, \citenamefont {Macri}, \citenamefont {Bowers}, \citenamefont {Scolnic}, \citenamefont {Cantat-Gaudin}, \citenamefont {Anderson},\ and\ \citenamefont {Reyes}}]{Riess_2022}%
  \BibitemOpen
  \bibfield  {author} {\bibinfo {author} {\bibfnamefont {A.~G.}\ \bibnamefont {Riess}}, \bibinfo {author} {\bibfnamefont {L.}~\bibnamefont {Breuval}}, \bibinfo {author} {\bibfnamefont {W.}~\bibnamefont {Yuan}}, \bibinfo {author} {\bibfnamefont {S.}~\bibnamefont {Casertano}}, \bibinfo {author} {\bibfnamefont {L.~M.}\ \bibnamefont {Macri}}, \bibinfo {author} {\bibfnamefont {J.~B.}\ \bibnamefont {Bowers}}, \bibinfo {author} {\bibfnamefont {D.}~\bibnamefont {Scolnic}}, \bibinfo {author} {\bibfnamefont {T.}~\bibnamefont {Cantat-Gaudin}}, \bibinfo {author} {\bibfnamefont {R.~I.}\ \bibnamefont {Anderson}}, \ and\ \bibinfo {author} {\bibfnamefont {M.~C.}\ \bibnamefont {Reyes}},\ }\href {\doibase 10.3847/1538-4357/ac8f24} {\bibfield  {journal} {\bibinfo  {journal} {The Astrophysical Journal}\ }\textbf {\bibinfo {volume} {938}},\ \bibinfo {pages} {36} (\bibinfo {year} {2022}{\natexlab{b}})}\BibitemShut {NoStop}%
\bibitem [{\citenamefont {Murakami}\ \emph {et~al.}(2023)\citenamefont {Murakami}, \citenamefont {Riess}, \citenamefont {Stahl}, \citenamefont {Kenworthy}, \citenamefont {Pluck}, \citenamefont {Macoretta}, \citenamefont {Brout}, \citenamefont {Jones}, \citenamefont {Scolnic},\ and\ \citenamefont {Filippenko}}]{murakami2023leveraging}%
  \BibitemOpen
  \bibfield  {author} {\bibinfo {author} {\bibfnamefont {Y.~S.}\ \bibnamefont {Murakami}}, \bibinfo {author} {\bibfnamefont {A.~G.}\ \bibnamefont {Riess}}, \bibinfo {author} {\bibfnamefont {B.~E.}\ \bibnamefont {Stahl}}, \bibinfo {author} {\bibfnamefont {W.~D.}\ \bibnamefont {Kenworthy}}, \bibinfo {author} {\bibfnamefont {D.-M.~A.}\ \bibnamefont {Pluck}}, \bibinfo {author} {\bibfnamefont {A.}~\bibnamefont {Macoretta}}, \bibinfo {author} {\bibfnamefont {D.}~\bibnamefont {Brout}}, \bibinfo {author} {\bibfnamefont {D.~O.}\ \bibnamefont {Jones}}, \bibinfo {author} {\bibfnamefont {D.~M.}\ \bibnamefont {Scolnic}}, \ and\ \bibinfo {author} {\bibfnamefont {A.~V.}\ \bibnamefont {Filippenko}},\ }\href@noop {} {\  (\bibinfo {year} {2023})},\ \Eprint {http://arxiv.org/abs/2306.00070} {arXiv:2306.00070 [astro-ph.CO]} \BibitemShut {NoStop}%
\bibitem [{\citenamefont {Di~Valentino}\ \emph {et~al.}(2021)\citenamefont {Di~Valentino}, \citenamefont {Mena}, \citenamefont {Pan}, \citenamefont {Visinelli}, \citenamefont {Yang}, \citenamefont {Melchiorri}, \citenamefont {Mota}, \citenamefont {Riess},\ and\ \citenamefont {Silk}}]{DiValentino:2021izs}%
  \BibitemOpen
  \bibfield  {author} {\bibinfo {author} {\bibfnamefont {E.}~\bibnamefont {Di~Valentino}}, \bibinfo {author} {\bibfnamefont {O.}~\bibnamefont {Mena}}, \bibinfo {author} {\bibfnamefont {S.}~\bibnamefont {Pan}}, \bibinfo {author} {\bibfnamefont {L.}~\bibnamefont {Visinelli}}, \bibinfo {author} {\bibfnamefont {W.}~\bibnamefont {Yang}}, \bibinfo {author} {\bibfnamefont {A.}~\bibnamefont {Melchiorri}}, \bibinfo {author} {\bibfnamefont {D.~F.}\ \bibnamefont {Mota}}, \bibinfo {author} {\bibfnamefont {A.~G.}\ \bibnamefont {Riess}}, \ and\ \bibinfo {author} {\bibfnamefont {J.}~\bibnamefont {Silk}},\ }\href {\doibase 10.1088/1361-6382/ac086d} {\bibfield  {journal} {\bibinfo  {journal} {Class. Quant. Grav.}\ }\textbf {\bibinfo {volume} {38}},\ \bibinfo {pages} {153001} (\bibinfo {year} {2021})},\ \Eprint {http://arxiv.org/abs/2103.01183} {arXiv:2103.01183 [astro-ph.CO]} \BibitemShut {NoStop}%
\bibitem [{\citenamefont {Collaboration}(2024)}]{desicollaboration2024desi}%
  \BibitemOpen
  \bibfield  {author} {\bibinfo {author} {\bibfnamefont {D.}~\bibnamefont {Collaboration}},\ }\href@noop {} {\enquote {\bibinfo {title} {Desi 2024 vi: Cosmological constraints from the measurements of baryon acoustic oscillations},}\ } (\bibinfo {year} {2024}),\ \Eprint {http://arxiv.org/abs/2404.03002} {arXiv:2404.03002 [astro-ph.CO]} \BibitemShut {NoStop}%
\bibitem [{\citenamefont {Alam}\ \emph {et~al.}(2021)\citenamefont {Alam} \emph {et~al.}}]{eBOSS:2020yzd}%
  \BibitemOpen
  \bibfield  {author} {\bibinfo {author} {\bibfnamefont {S.}~\bibnamefont {Alam}} \emph {et~al.} (\bibinfo {collaboration} {eBOSS}),\ }\href {\doibase 10.1103/PhysRevD.103.083533} {\bibfield  {journal} {\bibinfo  {journal} {Phys. Rev. D}\ }\textbf {\bibinfo {volume} {103}},\ \bibinfo {pages} {083533} (\bibinfo {year} {2021})},\ \Eprint {http://arxiv.org/abs/2007.08991} {arXiv:2007.08991 [astro-ph.CO]} \BibitemShut {NoStop}%
\bibitem [{\citenamefont {Choi}\ \emph {et~al.}(2020)\citenamefont {Choi} \emph {et~al.}}]{ACT:2020frw}%
  \BibitemOpen
  \bibfield  {author} {\bibinfo {author} {\bibfnamefont {S.~K.}\ \bibnamefont {Choi}} \emph {et~al.} (\bibinfo {collaboration} {ACT}),\ }\href {\doibase 10.1088/1475-7516/2020/12/045} {\bibfield  {journal} {\bibinfo  {journal} {JCAP}\ }\textbf {\bibinfo {volume} {12}},\ \bibinfo {pages} {045} (\bibinfo {year} {2020})},\ \Eprint {http://arxiv.org/abs/2007.07289} {arXiv:2007.07289 [astro-ph.CO]} \BibitemShut {NoStop}%
\bibitem [{\citenamefont {Dutcher}\ \emph {et~al.}(2021)\citenamefont {Dutcher}, \citenamefont {Balkenhol}, \citenamefont {Ade}, \citenamefont {Ahmed}, \citenamefont {Anderes}, \citenamefont {Anderson}, \citenamefont {Archipley} \emph {et~al.}}]{Dutcher:2021vtw}%
  \BibitemOpen
  \bibfield  {author} {\bibinfo {author} {\bibfnamefont {D.}~\bibnamefont {Dutcher}}, \bibinfo {author} {\bibfnamefont {L.}~\bibnamefont {Balkenhol}}, \bibinfo {author} {\bibfnamefont {P.~A.~R.}\ \bibnamefont {Ade}}, \bibinfo {author} {\bibfnamefont {Z.}~\bibnamefont {Ahmed}}, \bibinfo {author} {\bibfnamefont {E.}~\bibnamefont {Anderes}}, \bibinfo {author} {\bibfnamefont {A.~J.}\ \bibnamefont {Anderson}}, \bibinfo {author} {\bibnamefont {Archipley}},  \emph {et~al.} (\bibinfo {collaboration} {SPT-3G Collaboration}),\ }\href {\doibase 10.1103/PhysRevD.104.022003} {\bibfield  {journal} {\bibinfo  {journal} {Phys. Rev. D}\ }\textbf {\bibinfo {volume} {104}},\ \bibinfo {pages} {022003} (\bibinfo {year} {2021})}\BibitemShut {NoStop}%
\bibitem [{\citenamefont {Balkenhol}\ \emph {et~al.}(2023)\citenamefont {Balkenhol} \emph {et~al.}}]{SPT-3G:2022hvq}%
  \BibitemOpen
  \bibfield  {author} {\bibinfo {author} {\bibfnamefont {L.}~\bibnamefont {Balkenhol}} \emph {et~al.} (\bibinfo {collaboration} {SPT-3G}),\ }\href {\doibase 10.1103/PhysRevD.108.023510} {\bibfield  {journal} {\bibinfo  {journal} {Phys. Rev. D}\ }\textbf {\bibinfo {volume} {108}},\ \bibinfo {pages} {023510} (\bibinfo {year} {2023})},\ \Eprint {http://arxiv.org/abs/2212.05642} {arXiv:2212.05642 [astro-ph.CO]} \BibitemShut {NoStop}%
\bibitem [{\citenamefont {Riess}\ \emph {et~al.}(2024)\citenamefont {Riess}, \citenamefont {Anand}, \citenamefont {Yuan}, \citenamefont {Casertano}, \citenamefont {Dolphin}, \citenamefont {Macri}, \citenamefont {Breuval}, \citenamefont {Scolnic}, \citenamefont {Perrin},\ and\ \citenamefont {Anderson}}]{Riess:2024ohe}%
  \BibitemOpen
  \bibfield  {author} {\bibinfo {author} {\bibfnamefont {A.~G.}\ \bibnamefont {Riess}}, \bibinfo {author} {\bibfnamefont {G.~S.}\ \bibnamefont {Anand}}, \bibinfo {author} {\bibfnamefont {W.}~\bibnamefont {Yuan}}, \bibinfo {author} {\bibfnamefont {S.}~\bibnamefont {Casertano}}, \bibinfo {author} {\bibfnamefont {A.}~\bibnamefont {Dolphin}}, \bibinfo {author} {\bibfnamefont {L.~M.}\ \bibnamefont {Macri}}, \bibinfo {author} {\bibfnamefont {L.}~\bibnamefont {Breuval}}, \bibinfo {author} {\bibfnamefont {D.}~\bibnamefont {Scolnic}}, \bibinfo {author} {\bibfnamefont {M.}~\bibnamefont {Perrin}}, \ and\ \bibinfo {author} {\bibfnamefont {I.~R.}\ \bibnamefont {Anderson}},\ }\href {\doibase 10.3847/2041-8213/ad1ddd} {\bibfield  {journal} {\bibinfo  {journal} {Astrophys. J. Lett.}\ }\textbf {\bibinfo {volume} {962}},\ \bibinfo {pages} {L17} (\bibinfo {year} {2024})},\ \Eprint {http://arxiv.org/abs/2401.04773} {arXiv:2401.04773 [astro-ph.CO]} \BibitemShut {NoStop}%
\bibitem [{\citenamefont {Abdalla}\ \emph {et~al.}(2022)\citenamefont {Abdalla} \emph {et~al.}}]{Abdalla:2022yfr}%
  \BibitemOpen
  \bibfield  {author} {\bibinfo {author} {\bibfnamefont {E.}~\bibnamefont {Abdalla}} \emph {et~al.},\ }\href {\doibase 10.1016/j.jheap.2022.04.002} {\bibfield  {journal} {\bibinfo  {journal} {JHEAp}\ }\textbf {\bibinfo {volume} {34}},\ \bibinfo {pages} {49} (\bibinfo {year} {2022})},\ \Eprint {http://arxiv.org/abs/2203.06142} {arXiv:2203.06142 [astro-ph.CO]} \BibitemShut {NoStop}%
\bibitem [{\citenamefont {Perivolaropoulos}\ and\ \citenamefont {Skara}(2022)}]{Perivolaropoulos_2022}%
  \BibitemOpen
  \bibfield  {author} {\bibinfo {author} {\bibfnamefont {L.}~\bibnamefont {Perivolaropoulos}}\ and\ \bibinfo {author} {\bibfnamefont {F.}~\bibnamefont {Skara}},\ }\href {\doibase 10.1016/j.newar.2022.101659} {\bibfield  {journal} {\bibinfo  {journal} {New Astronomy Reviews}\ }\textbf {\bibinfo {volume} {95}},\ \bibinfo {pages} {101659} (\bibinfo {year} {2022})}\BibitemShut {NoStop}%
\bibitem [{\citenamefont {{Holanda}}\ \emph {et~al.}(2012{\natexlab{a}})\citenamefont {{Holanda}}, \citenamefont {{Cunha}}, \citenamefont {{Marassi}},\ and\ \citenamefont {{Lima}}}]{2012JCAP...02..035H}%
  \BibitemOpen
  \bibfield  {author} {\bibinfo {author} {\bibfnamefont {R.~F.~L.}\ \bibnamefont {{Holanda}}}, \bibinfo {author} {\bibfnamefont {J.~V.}\ \bibnamefont {{Cunha}}}, \bibinfo {author} {\bibfnamefont {L.}~\bibnamefont {{Marassi}}}, \ and\ \bibinfo {author} {\bibfnamefont {J.~A.~S.}\ \bibnamefont {{Lima}}},\ }\href {\doibase 10.1088/1475-7516/2012/02/035} {\bibfield  {journal} {\bibinfo  {journal} {jcap}\ }\textbf {\bibinfo {volume} {2012}},\ \bibinfo {eid} {035} (\bibinfo {year} {2012}{\natexlab{a}})},\ \Eprint {http://arxiv.org/abs/1006.4200} {arXiv:1006.4200 [astro-ph.CO]} \BibitemShut {NoStop}%
\bibitem [{\citenamefont {{Holanda}}\ \emph {et~al.}(2014)\citenamefont {{Holanda}}, \citenamefont {{Busti}},\ and\ \citenamefont {{Pordeus da Silva}}}]{2014MNRAS.443L..74H}%
  \BibitemOpen
  \bibfield  {author} {\bibinfo {author} {\bibfnamefont {R.~F.~L.}\ \bibnamefont {{Holanda}}}, \bibinfo {author} {\bibfnamefont {V.~C.}\ \bibnamefont {{Busti}}}, \ and\ \bibinfo {author} {\bibfnamefont {G.}~\bibnamefont {{Pordeus da Silva}}},\ }\href {\doibase 10.1093/mnrasl/slu086} {\bibfield  {journal} {\bibinfo  {journal} {mnras}\ }\textbf {\bibinfo {volume} {443}},\ \bibinfo {pages} {L74} (\bibinfo {year} {2014})},\ \Eprint {http://arxiv.org/abs/1404.4418} {arXiv:1404.4418 [astro-ph.CO]} \BibitemShut {NoStop}%
\bibitem [{\citenamefont {{Holanda}}\ \emph {et~al.}(2020)\citenamefont {{Holanda}}, \citenamefont {{Pordeus-da-Silva}},\ and\ \citenamefont {{Pereira}}}]{2020JCAP...09..053H}%
  \BibitemOpen
  \bibfield  {author} {\bibinfo {author} {\bibfnamefont {R.~F.~L.}\ \bibnamefont {{Holanda}}}, \bibinfo {author} {\bibfnamefont {G.}~\bibnamefont {{Pordeus-da-Silva}}}, \ and\ \bibinfo {author} {\bibfnamefont {S.~H.}\ \bibnamefont {{Pereira}}},\ }\href {\doibase 10.1088/1475-7516/2020/09/053} {\bibfield  {journal} {\bibinfo  {journal} {jcap}\ }\textbf {\bibinfo {volume} {2020}},\ \bibinfo {eid} {053} (\bibinfo {year} {2020})},\ \Eprint {http://arxiv.org/abs/2006.06712} {arXiv:2006.06712 [astro-ph.CO]} \BibitemShut {NoStop}%
\bibitem [{\citenamefont {Mantz}\ \emph {et~al.}(2021)\citenamefont {Mantz}, \citenamefont {Morris}, \citenamefont {Allen}, \citenamefont {Canning}, \citenamefont {Baumont}, \citenamefont {Benson}, \citenamefont {Bleem}, \citenamefont {Ehlert}, \citenamefont {Floyd}, \citenamefont {Herbonnet}, \citenamefont {Kelly}, \citenamefont {Liang}, \citenamefont {von der Linden}, \citenamefont {McDonald}, \citenamefont {Rapetti}, \citenamefont {Schmidt}, \citenamefont {Werner},\ and\ \citenamefont {Wright}}]{Mantz_2021}%
  \BibitemOpen
  \bibfield  {author} {\bibinfo {author} {\bibfnamefont {A.~B.}\ \bibnamefont {Mantz}}, \bibinfo {author} {\bibfnamefont {R.~G.}\ \bibnamefont {Morris}}, \bibinfo {author} {\bibfnamefont {S.~W.}\ \bibnamefont {Allen}}, \bibinfo {author} {\bibfnamefont {R.~E.~A.}\ \bibnamefont {Canning}}, \bibinfo {author} {\bibfnamefont {L.}~\bibnamefont {Baumont}}, \bibinfo {author} {\bibfnamefont {B.}~\bibnamefont {Benson}}, \bibinfo {author} {\bibfnamefont {L.~E.}\ \bibnamefont {Bleem}}, \bibinfo {author} {\bibfnamefont {S.~R.}\ \bibnamefont {Ehlert}}, \bibinfo {author} {\bibfnamefont {B.}~\bibnamefont {Floyd}}, \bibinfo {author} {\bibfnamefont {R.}~\bibnamefont {Herbonnet}}, \bibinfo {author} {\bibfnamefont {P.~L.}\ \bibnamefont {Kelly}}, \bibinfo {author} {\bibfnamefont {S.}~\bibnamefont {Liang}}, \bibinfo {author} {\bibfnamefont {A.}~\bibnamefont {von der Linden}}, \bibinfo {author} {\bibfnamefont {M.}~\bibnamefont {McDonald}}, \bibinfo {author} {\bibfnamefont {D.~A.}\ \bibnamefont {Rapetti}}, \bibinfo {author}
  {\bibfnamefont {R.~W.}\ \bibnamefont {Schmidt}}, \bibinfo {author} {\bibfnamefont {N.}~\bibnamefont {Werner}}, \ and\ \bibinfo {author} {\bibfnamefont {A.}~\bibnamefont {Wright}},\ }\href {\doibase 10.1093/mnras/stab3390} {\bibfield  {journal} {\bibinfo  {journal} {Monthly Notices of the Royal Astronomical Society}\ }\textbf {\bibinfo {volume} {510}},\ \bibinfo {pages} {131–145} (\bibinfo {year} {2021})}\BibitemShut {NoStop}%
\bibitem [{\citenamefont {{Holanda}}\ \emph {et~al.}(2012{\natexlab{b}})\citenamefont {{Holanda}}, \citenamefont {{Cunha}},\ and\ \citenamefont {{Lima}}}]{2012GReGr..44..501H}%
  \BibitemOpen
  \bibfield  {author} {\bibinfo {author} {\bibfnamefont {R.~F.~L.}\ \bibnamefont {{Holanda}}}, \bibinfo {author} {\bibfnamefont {J.~V.}\ \bibnamefont {{Cunha}}}, \ and\ \bibinfo {author} {\bibfnamefont {J.~A.~S.}\ \bibnamefont {{Lima}}},\ }\href {\doibase 10.1007/s10714-011-1292-5} {\bibfield  {journal} {\bibinfo  {journal} {General Relativity and Gravitation}\ }\textbf {\bibinfo {volume} {44}},\ \bibinfo {pages} {501} (\bibinfo {year} {2012}{\natexlab{b}})},\ \Eprint {http://arxiv.org/abs/0807.0647} {arXiv:0807.0647 [astro-ph]} \BibitemShut {NoStop}%
\bibitem [{\citenamefont {{Cunha}}\ \emph {et~al.}(2007)\citenamefont {{Cunha}}, \citenamefont {{Marassi}},\ and\ \citenamefont {{Lima}}}]{2007MNRAS.379L...1C}%
  \BibitemOpen
  \bibfield  {author} {\bibinfo {author} {\bibfnamefont {J.~V.}\ \bibnamefont {{Cunha}}}, \bibinfo {author} {\bibfnamefont {L.}~\bibnamefont {{Marassi}}}, \ and\ \bibinfo {author} {\bibfnamefont {J.~A.~S.}\ \bibnamefont {{Lima}}},\ }\href {\doibase 10.1111/j.1745-3933.2007.00322.x} {\bibfield  {journal} {\bibinfo  {journal} {mnras}\ }\textbf {\bibinfo {volume} {379}},\ \bibinfo {pages} {L1} (\bibinfo {year} {2007})},\ \Eprint {http://arxiv.org/abs/astro-ph/0611934} {arXiv:astro-ph/0611934 [astro-ph]} \BibitemShut {NoStop}%
\bibitem [{\citenamefont {{Bora}}\ and\ \citenamefont {{Holanda}}(2023)}]{2023EPJC...83..274B}%
  \BibitemOpen
  \bibfield  {author} {\bibinfo {author} {\bibfnamefont {K.}~\bibnamefont {{Bora}}}\ and\ \bibinfo {author} {\bibfnamefont {R.~F.~L.}\ \bibnamefont {{Holanda}}},\ }\href {\doibase 10.1140/epjc/s10052-023-11424-y} {\bibfield  {journal} {\bibinfo  {journal} {European Physical Journal C}\ }\textbf {\bibinfo {volume} {83}},\ \bibinfo {eid} {274} (\bibinfo {year} {2023})},\ \Eprint {http://arxiv.org/abs/2203.07223} {arXiv:2203.07223 [astro-ph.CO]} \BibitemShut {NoStop}%
\bibitem [{\citenamefont {{Kozmanyan}}\ \emph {et~al.}(2019)\citenamefont {{Kozmanyan}}, \citenamefont {{Bourdin}}, \citenamefont {{Mazzotta}}, \citenamefont {{Rasia}},\ and\ \citenamefont {{Sereno}}}]{2019A&A...621A..34K}%
  \BibitemOpen
  \bibfield  {author} {\bibinfo {author} {\bibfnamefont {A.}~\bibnamefont {{Kozmanyan}}}, \bibinfo {author} {\bibfnamefont {H.}~\bibnamefont {{Bourdin}}}, \bibinfo {author} {\bibfnamefont {P.}~\bibnamefont {{Mazzotta}}}, \bibinfo {author} {\bibfnamefont {E.}~\bibnamefont {{Rasia}}}, \ and\ \bibinfo {author} {\bibfnamefont {M.}~\bibnamefont {{Sereno}}},\ }\href {\doibase 10.1051/0004-6361/201833879} {\bibfield  {journal} {\bibinfo  {journal} {aap}\ }\textbf {\bibinfo {volume} {621}},\ \bibinfo {eid} {A34} (\bibinfo {year} {2019})},\ \Eprint {http://arxiv.org/abs/1809.09560} {arXiv:1809.09560 [astro-ph.CO]} \BibitemShut {NoStop}%
\bibitem [{\citenamefont {Barbosa}\ \emph {et~al.}(2024)\citenamefont {Barbosa}, \citenamefont {von Marttens}, \citenamefont {Gonzalez},\ and\ \citenamefont {Alcaniz}}]{barbosa2024assessing}%
  \BibitemOpen
  \bibfield  {author} {\bibinfo {author} {\bibfnamefont {D.}~\bibnamefont {Barbosa}}, \bibinfo {author} {\bibfnamefont {R.}~\bibnamefont {von Marttens}}, \bibinfo {author} {\bibfnamefont {J.}~\bibnamefont {Gonzalez}}, \ and\ \bibinfo {author} {\bibfnamefont {J.}~\bibnamefont {Alcaniz}},\ }\href@noop {} {\enquote {\bibinfo {title} {Assessing the dark degeneracy through the gas mass fraction data},}\ } (\bibinfo {year} {2024}),\ \Eprint {http://arxiv.org/abs/2403.12220} {arXiv:2403.12220 [astro-ph.CO]} \BibitemShut {NoStop}%
\bibitem [{\citenamefont {Panchal}\ and\ \citenamefont {Desai}(2024)}]{panchal2024comparison}%
  \BibitemOpen
  \bibfield  {author} {\bibinfo {author} {\bibfnamefont {K.}~\bibnamefont {Panchal}}\ and\ \bibinfo {author} {\bibfnamefont {S.}~\bibnamefont {Desai}},\ }\href@noop {} {\enquote {\bibinfo {title} {Comparison of $\lambda$cdm and $r_h = ct$ with updated galaxy cluster $f_{gas}$ measurements using bayesian inference},}\ } (\bibinfo {year} {2024}),\ \Eprint {http://arxiv.org/abs/2401.11138} {arXiv:2401.11138 [astro-ph.CO]} \BibitemShut {NoStop}%
\bibitem [{\citenamefont {Darragh-Ford}\ \emph {et~al.}(2023)\citenamefont {Darragh-Ford}, \citenamefont {Mantz}, \citenamefont {Rasia}, \citenamefont {Allen}, \citenamefont {Morris}, \citenamefont {Foster}, \citenamefont {Schmidt},\ and\ \citenamefont {Wenrich}}]{Darragh_Ford_2023}%
  \BibitemOpen
  \bibfield  {author} {\bibinfo {author} {\bibfnamefont {E.}~\bibnamefont {Darragh-Ford}}, \bibinfo {author} {\bibfnamefont {A.~B.}\ \bibnamefont {Mantz}}, \bibinfo {author} {\bibfnamefont {E.}~\bibnamefont {Rasia}}, \bibinfo {author} {\bibfnamefont {S.~W.}\ \bibnamefont {Allen}}, \bibinfo {author} {\bibfnamefont {R.~G.}\ \bibnamefont {Morris}}, \bibinfo {author} {\bibfnamefont {J.}~\bibnamefont {Foster}}, \bibinfo {author} {\bibfnamefont {R.~W.}\ \bibnamefont {Schmidt}}, \ and\ \bibinfo {author} {\bibfnamefont {G.}~\bibnamefont {Wenrich}},\ }\href {\doibase 10.1093/mnras/stad585} {\bibfield  {journal} {\bibinfo  {journal} {Monthly Notices of the Royal Astronomical Society}\ }\textbf {\bibinfo {volume} {521}},\ \bibinfo {pages} {790–799} (\bibinfo {year} {2023})}\BibitemShut {NoStop}%
\bibitem [{\citenamefont {Wicker}\ \emph {et~al.}(2023)\citenamefont {Wicker}, \citenamefont {Douspis}, \citenamefont {Salvati},\ and\ \citenamefont {Aghanim}}]{Wicker_2023}%
  \BibitemOpen
  \bibfield  {author} {\bibinfo {author} {\bibfnamefont {R.}~\bibnamefont {Wicker}}, \bibinfo {author} {\bibfnamefont {M.}~\bibnamefont {Douspis}}, \bibinfo {author} {\bibfnamefont {L.}~\bibnamefont {Salvati}}, \ and\ \bibinfo {author} {\bibfnamefont {N.}~\bibnamefont {Aghanim}},\ }\href {\doibase 10.1051/0004-6361/202243922} {\bibfield  {journal} {\bibinfo  {journal} {Astronomy \& Astrophysics}\ }\textbf {\bibinfo {volume} {674}},\ \bibinfo {pages} {A48} (\bibinfo {year} {2023})}\BibitemShut {NoStop}%
\bibitem [{\citenamefont {Cola\c{c}o}\ \emph {et~al.}(2024)\citenamefont {Cola\c{c}o}, \citenamefont {Ferreira}, \citenamefont {Holanda}, \citenamefont {Gonzalez},\ and\ \citenamefont {Nunes}}]{Colaco:2023gzy}%
  \BibitemOpen
  \bibfield  {author} {\bibinfo {author} {\bibfnamefont {L.~R.}\ \bibnamefont {Cola\c{c}o}}, \bibinfo {author} {\bibfnamefont {M.}~\bibnamefont {Ferreira}}, \bibinfo {author} {\bibfnamefont {R.~F.~L.}\ \bibnamefont {Holanda}}, \bibinfo {author} {\bibfnamefont {J.~E.}\ \bibnamefont {Gonzalez}}, \ and\ \bibinfo {author} {\bibfnamefont {R.~C.}\ \bibnamefont {Nunes}},\ }\href {\doibase 10.1088/1475-7516/2024/05/098} {\bibfield  {journal} {\bibinfo  {journal} {JCAP}\ }\textbf {\bibinfo {volume} {05}},\ \bibinfo {pages} {098} (\bibinfo {year} {2024})},\ \Eprint {http://arxiv.org/abs/2310.18711} {arXiv:2310.18711 [astro-ph.CO]} \BibitemShut {NoStop}%
\bibitem [{\citenamefont {{Etherington}}(1933)}]{1933PMag...15..761E}%
  \BibitemOpen
  \bibfield  {author} {\bibinfo {author} {\bibfnamefont {I.~M.~H.}\ \bibnamefont {{Etherington}}},\ }\href@noop {} {\bibfield  {journal} {\bibinfo  {journal} {Philosophical Magazine}\ }\textbf {\bibinfo {volume} {15}},\ \bibinfo {pages} {761} (\bibinfo {year} {1933})}\BibitemShut {NoStop}%
\bibitem [{\citenamefont {Renzi}\ and\ \citenamefont {Silvestri}(2023)}]{Renzi:2020fnx}%
  \BibitemOpen
  \bibfield  {author} {\bibinfo {author} {\bibfnamefont {F.}~\bibnamefont {Renzi}}\ and\ \bibinfo {author} {\bibfnamefont {A.}~\bibnamefont {Silvestri}},\ }\href {\doibase 10.1103/PhysRevD.107.023520} {\bibfield  {journal} {\bibinfo  {journal} {Phys. Rev. D}\ }\textbf {\bibinfo {volume} {107}},\ \bibinfo {pages} {023520} (\bibinfo {year} {2023})},\ \Eprint {http://arxiv.org/abs/2011.10559} {arXiv:2011.10559 [astro-ph.CO]} \BibitemShut {NoStop}%
\bibitem [{\citenamefont {{Krolewski}}\ and\ \citenamefont {{Percival}}(2024)}]{2024arXiv240319236K}%
  \BibitemOpen
  \bibfield  {author} {\bibinfo {author} {\bibfnamefont {A.}~\bibnamefont {{Krolewski}}}\ and\ \bibinfo {author} {\bibfnamefont {W.~J.}\ \bibnamefont {{Percival}}},\ }\href {\doibase 10.48550/arXiv.2403.19236} {\bibfield  {journal} {\bibinfo  {journal} {arXiv e-prints}\ ,\ \bibinfo {eid} {arXiv:2403.19236}} (\bibinfo {year} {2024})},\ \Eprint {http://arxiv.org/abs/2403.19236} {arXiv:2403.19236 [astro-ph.CO]} \BibitemShut {NoStop}%
\bibitem [{\citenamefont {Eckert}\ \emph {et~al.}()\citenamefont {Eckert}, \citenamefont {Ghirardini}, \citenamefont {Ettori}, \citenamefont {Rasia}, \citenamefont {Biffi}, \citenamefont {Pointecouteau}, \citenamefont {Rossetti}, \citenamefont {Molendi}, \citenamefont {Vazza}, \citenamefont {Gastaldello} \emph {et~al.}}]{eckert2019non}%
  \BibitemOpen
  \bibfield  {author} {\bibinfo {author} {\bibfnamefont {D.}~\bibnamefont {Eckert}}, \bibinfo {author} {\bibfnamefont {V.}~\bibnamefont {Ghirardini}}, \bibinfo {author} {\bibfnamefont {S.}~\bibnamefont {Ettori}}, \bibinfo {author} {\bibfnamefont {E.}~\bibnamefont {Rasia}}, \bibinfo {author} {\bibfnamefont {V.}~\bibnamefont {Biffi}}, \bibinfo {author} {\bibfnamefont {E.}~\bibnamefont {Pointecouteau}}, \bibinfo {author} {\bibfnamefont {M.}~\bibnamefont {Rossetti}}, \bibinfo {author} {\bibfnamefont {S.}~\bibnamefont {Molendi}}, \bibinfo {author} {\bibfnamefont {F.}~\bibnamefont {Vazza}}, \bibinfo {author} {\bibfnamefont {F.}~\bibnamefont {Gastaldello}},  \emph {et~al.},\ }\href@noop {} {\ }\bibinfo {note} {\textit{Astron. Astrophys.} \textbf{621}, A40 (2019)}\BibitemShut {NoStop}%
\bibitem [{\citenamefont {Ettori}\ \emph {et~al.}()\citenamefont {Ettori}, \citenamefont {Gastaldello}, \citenamefont {Leccardi}, \citenamefont {Molendi}, \citenamefont {Rossetti}, \citenamefont {Buote},\ and\ \citenamefont {Meneghetti}}]{ettori2010mass}%
  \BibitemOpen
  \bibfield  {author} {\bibinfo {author} {\bibfnamefont {S.}~\bibnamefont {Ettori}}, \bibinfo {author} {\bibfnamefont {F.}~\bibnamefont {Gastaldello}}, \bibinfo {author} {\bibfnamefont {A.}~\bibnamefont {Leccardi}}, \bibinfo {author} {\bibfnamefont {S.}~\bibnamefont {Molendi}}, \bibinfo {author} {\bibfnamefont {M.}~\bibnamefont {Rossetti}}, \bibinfo {author} {\bibfnamefont {D.}~\bibnamefont {Buote}}, \ and\ \bibinfo {author} {\bibfnamefont {M.}~\bibnamefont {Meneghetti}},\ }\href@noop {} {\ }\bibinfo {note} {\textit{Astron. Astrophys.} \textbf{524}, A68 (2010)}\BibitemShut {NoStop}%
\bibitem [{\citenamefont {Ghirardini}\ \emph {et~al.}()\citenamefont {Ghirardini}, \citenamefont {Ettori}, \citenamefont {Amodeo}, \citenamefont {Capasso},\ and\ \citenamefont {Sereno}}]{ghirardini2017evolution}%
  \BibitemOpen
  \bibfield  {author} {\bibinfo {author} {\bibfnamefont {V.}~\bibnamefont {Ghirardini}}, \bibinfo {author} {\bibfnamefont {S.}~\bibnamefont {Ettori}}, \bibinfo {author} {\bibfnamefont {S.}~\bibnamefont {Amodeo}}, \bibinfo {author} {\bibfnamefont {R.}~\bibnamefont {Capasso}}, \ and\ \bibinfo {author} {\bibfnamefont {M.}~\bibnamefont {Sereno}},\ }\href@noop {} {\ }\bibinfo {note} {\textit{Astron. Astrophys.} \textbf{604}, A100 (2017)}\BibitemShut {NoStop}%
\bibitem [{\citenamefont {Corasaniti}\ \emph {et~al.}()\citenamefont {Corasaniti}, \citenamefont {Sereno},\ and\ \citenamefont {Ettori}}]{corasaniti2021cosmological}%
  \BibitemOpen
  \bibfield  {author} {\bibinfo {author} {\bibfnamefont {P.-S.}\ \bibnamefont {Corasaniti}}, \bibinfo {author} {\bibfnamefont {M.}~\bibnamefont {Sereno}}, \ and\ \bibinfo {author} {\bibfnamefont {S.}~\bibnamefont {Ettori}},\ }\href@noop {} {\ }\bibinfo {note} {\textit{Astrophys. J.} \textbf{911}, 82 (2021)}\BibitemShut {NoStop}%
\bibitem [{\citenamefont {Scolnic}\ \emph {et~al.}(2018)\citenamefont {Scolnic} \emph {et~al.}}]{Scolnic:2017caz}%
  \BibitemOpen
  \bibfield  {author} {\bibinfo {author} {\bibfnamefont {D.~M.}\ \bibnamefont {Scolnic}} \emph {et~al.},\ }\href {\doibase 10.3847/1538-4357/aab9bb} {\bibfield  {journal} {\bibinfo  {journal} {Astrophys. J.}\ }\textbf {\bibinfo {volume} {859}},\ \bibinfo {pages} {101} (\bibinfo {year} {2018})},\ \Eprint {http://arxiv.org/abs/1710.00845} {arXiv:1710.00845 [astro-ph.CO]} \BibitemShut {NoStop}%
\bibitem [{\citenamefont {{Holanda}}\ \emph {et~al.}(2012{\natexlab{c}})\citenamefont {{Holanda}}, \citenamefont {{Lima}},\ and\ \citenamefont {{Ribeiro}}}]{2012A&A...538A.131H}%
  \BibitemOpen
  \bibfield  {author} {\bibinfo {author} {\bibfnamefont {R.~F.~L.}\ \bibnamefont {{Holanda}}}, \bibinfo {author} {\bibfnamefont {J.~A.~S.}\ \bibnamefont {{Lima}}}, \ and\ \bibinfo {author} {\bibfnamefont {M.~B.}\ \bibnamefont {{Ribeiro}}},\ }\href {\doibase 10.1051/0004-6361/201118343} {\bibfield  {journal} {\bibinfo  {journal} {aap}\ }\textbf {\bibinfo {volume} {538}},\ \bibinfo {eid} {A131} (\bibinfo {year} {2012}{\natexlab{c}})},\ \Eprint {http://arxiv.org/abs/1104.3753} {arXiv:1104.3753 [astro-ph.CO]} \BibitemShut {NoStop}%
\bibitem [{\citenamefont {{Xu}}\ \emph {et~al.}(2022)\citenamefont {{Xu}}, \citenamefont {{Wang}}, \citenamefont {{Zhang}}, \citenamefont {{Huang}},\ and\ \citenamefont {{Zhang}}}]{2022ApJ...939..115X}%
  \BibitemOpen
  \bibfield  {author} {\bibinfo {author} {\bibfnamefont {B.}~\bibnamefont {{Xu}}}, \bibinfo {author} {\bibfnamefont {Z.}~\bibnamefont {{Wang}}}, \bibinfo {author} {\bibfnamefont {K.}~\bibnamefont {{Zhang}}}, \bibinfo {author} {\bibfnamefont {Q.}~\bibnamefont {{Huang}}}, \ and\ \bibinfo {author} {\bibfnamefont {J.}~\bibnamefont {{Zhang}}},\ }\href {\doibase 10.3847/1538-4357/ac9793} {\bibfield  {journal} {\bibinfo  {journal} {apj}\ }\textbf {\bibinfo {volume} {939}},\ \bibinfo {eid} {115} (\bibinfo {year} {2022})},\ \Eprint {http://arxiv.org/abs/2212.00269} {arXiv:2212.00269 [astro-ph.CO]} \BibitemShut {NoStop}%
\bibitem [{\citenamefont {{Liao}}(2019)}]{2019ApJ...885...70L}%
  \BibitemOpen
  \bibfield  {author} {\bibinfo {author} {\bibfnamefont {K.}~\bibnamefont {{Liao}}},\ }\href {\doibase 10.3847/1538-4357/ab4819} {\bibfield  {journal} {\bibinfo  {journal} {apj}\ }\textbf {\bibinfo {volume} {885}},\ \bibinfo {eid} {70} (\bibinfo {year} {2019})},\ \Eprint {http://arxiv.org/abs/1906.09588} {arXiv:1906.09588 [astro-ph.CO]} \BibitemShut {NoStop}%
\bibitem [{\citenamefont {{Liu}}\ \emph {et~al.}(2021)\citenamefont {{Liu}}, \citenamefont {{Cao}}, \citenamefont {{Zhang}}, \citenamefont {{Gong}}, \citenamefont {{Guo}},\ and\ \citenamefont {{Zheng}}}]{2021EPJC...81..903L}%
  \BibitemOpen
  \bibfield  {author} {\bibinfo {author} {\bibfnamefont {T.}~\bibnamefont {{Liu}}}, \bibinfo {author} {\bibfnamefont {S.}~\bibnamefont {{Cao}}}, \bibinfo {author} {\bibfnamefont {S.}~\bibnamefont {{Zhang}}}, \bibinfo {author} {\bibfnamefont {X.}~\bibnamefont {{Gong}}}, \bibinfo {author} {\bibfnamefont {W.}~\bibnamefont {{Guo}}}, \ and\ \bibinfo {author} {\bibfnamefont {C.}~\bibnamefont {{Zheng}}},\ }\href {\doibase 10.1140/epjc/s10052-021-09713-5} {\bibfield  {journal} {\bibinfo  {journal} {European Physical Journal C}\ }\textbf {\bibinfo {volume} {81}},\ \bibinfo {eid} {903} (\bibinfo {year} {2021})},\ \Eprint {http://arxiv.org/abs/2110.00927} {arXiv:2110.00927 [astro-ph.CO]} \BibitemShut {NoStop}%
\bibitem [{\citenamefont {Etherington}(2007)}]{CDDR}%
  \BibitemOpen
  \bibfield  {author} {\bibinfo {author} {\bibfnamefont {I.~M.~H.}\ \bibnamefont {Etherington}},\ }\href@noop {} {\bibfield  {journal} {\bibinfo  {journal} {General Relativity and Gravitation}\ }\textbf {\bibinfo {volume} {39}} (\bibinfo {year} {2007})}\BibitemShut {NoStop}%
\bibitem [{\citenamefont {Bassett}\ and\ \citenamefont {Kunz}(2004)}]{Bassett:2003vu}%
  \BibitemOpen
  \bibfield  {author} {\bibinfo {author} {\bibfnamefont {B.~A.}\ \bibnamefont {Bassett}}\ and\ \bibinfo {author} {\bibfnamefont {M.}~\bibnamefont {Kunz}},\ }\href {\doibase 10.1103/PhysRevD.69.101305} {\bibfield  {journal} {\bibinfo  {journal} {Phys. Rev. D}\ }\textbf {\bibinfo {volume} {69}},\ \bibinfo {pages} {101305} (\bibinfo {year} {2004})},\ \Eprint {http://arxiv.org/abs/astro-ph/0312443} {arXiv:astro-ph/0312443} \BibitemShut {NoStop}%
\bibitem [{\citenamefont {Ellis}(2007)}]{Ellis2007}%
  \BibitemOpen
  \bibfield  {author} {\bibinfo {author} {\bibfnamefont {G.~F.~R.}\ \bibnamefont {Ellis}},\ }\href@noop {} {\bibfield  {journal} {\bibinfo  {journal} {General Relativity and Gravitation}\ }\textbf {\bibinfo {volume} {39}} (\bibinfo {year} {2007})}\BibitemShut {NoStop}%
\bibitem [{\citenamefont {{Sarazin}}(1986)}]{1986RvMP...58....1S}%
  \BibitemOpen
  \bibfield  {author} {\bibinfo {author} {\bibfnamefont {C.~L.}\ \bibnamefont {{Sarazin}}},\ }\href {\doibase 10.1103/RevModPhys.58.1} {\bibfield  {journal} {\bibinfo  {journal} {Reviews of Modern Physics}\ }\textbf {\bibinfo {volume} {58}},\ \bibinfo {pages} {1} (\bibinfo {year} {1986})}\BibitemShut {NoStop}%
\bibitem [{\citenamefont {Sarazin}(1986)}]{sarazin1986x}%
  \BibitemOpen
  \bibfield  {author} {\bibinfo {author} {\bibfnamefont {C.~L.}\ \bibnamefont {Sarazin}},\ }\href@noop {} {\bibfield  {journal} {\bibinfo  {journal} {Reviews of Modern Physics}\ }\textbf {\bibinfo {volume} {58}},\ \bibinfo {pages} {1} (\bibinfo {year} {1986})}\BibitemShut {NoStop}%
\bibitem [{\citenamefont {Cavaliere}\ and\ \citenamefont {Fusco-Femiano}(1978)}]{cavaliere1978distribution}%
  \BibitemOpen
  \bibfield  {author} {\bibinfo {author} {\bibfnamefont {A.}~\bibnamefont {Cavaliere}}\ and\ \bibinfo {author} {\bibfnamefont {R.}~\bibnamefont {Fusco-Femiano}},\ }\href@noop {} {\bibfield  {journal} {\bibinfo  {journal} {Astronomy and Astrophysics, Vol. 70, p. 677 (1978)}\ }\textbf {\bibinfo {volume} {70}},\ \bibinfo {pages} {677} (\bibinfo {year} {1978})}\BibitemShut {NoStop}%
\bibitem [{\citenamefont {{Allen}}\ \emph {et~al.}(2011{\natexlab{a}})\citenamefont {{Allen}}, \citenamefont {{Evrard}},\ and\ \citenamefont {{Mantz}}}]{2011ARAA..49..409A}%
  \BibitemOpen
  \bibfield  {author} {\bibinfo {author} {\bibfnamefont {S.~W.}\ \bibnamefont {{Allen}}}, \bibinfo {author} {\bibfnamefont {A.~E.}\ \bibnamefont {{Evrard}}}, \ and\ \bibinfo {author} {\bibfnamefont {A.~B.}\ \bibnamefont {{Mantz}}},\ }\href {\doibase 10.1146/annurev-astro-081710-102514} {\bibfield  {journal} {\bibinfo  {journal} {araa}\ }\textbf {\bibinfo {volume} {49}},\ \bibinfo {pages} {409} (\bibinfo {year} {2011}{\natexlab{a}})},\ \Eprint {http://arxiv.org/abs/1103.4829} {arXiv:1103.4829 [astro-ph.CO]} \BibitemShut {NoStop}%
\bibitem [{\citenamefont {{Allen}}\ \emph {et~al.}(2008)\citenamefont {{Allen}}, \citenamefont {{Rapetti}}, \citenamefont {{Schmidt}}, \citenamefont {{Ebeling}}, \citenamefont {{Morris}},\ and\ \citenamefont {{Fabian}}}]{2008MNRAS.383..879A}%
  \BibitemOpen
  \bibfield  {author} {\bibinfo {author} {\bibfnamefont {S.~W.}\ \bibnamefont {{Allen}}}, \bibinfo {author} {\bibfnamefont {D.~A.}\ \bibnamefont {{Rapetti}}}, \bibinfo {author} {\bibfnamefont {R.~W.}\ \bibnamefont {{Schmidt}}}, \bibinfo {author} {\bibfnamefont {H.}~\bibnamefont {{Ebeling}}}, \bibinfo {author} {\bibfnamefont {R.~G.}\ \bibnamefont {{Morris}}}, \ and\ \bibinfo {author} {\bibfnamefont {A.~C.}\ \bibnamefont {{Fabian}}},\ }\href {\doibase 10.1111/j.1365-2966.2007.12610.x} {\bibfield  {journal} {\bibinfo  {journal} {mnras}\ }\textbf {\bibinfo {volume} {383}},\ \bibinfo {pages} {879} (\bibinfo {year} {2008})},\ \Eprint {http://arxiv.org/abs/0706.0033} {arXiv:0706.0033 [astro-ph]} \BibitemShut {NoStop}%
\bibitem [{\citenamefont {{Ettori}}\ \emph {et~al.}(2009)\citenamefont {{Ettori}}, \citenamefont {{Morandi}}, \citenamefont {{Tozzi}}, \citenamefont {{Balestra}}, \citenamefont {{Borgani}}, \citenamefont {{Rosati}}, \citenamefont {{Lovisari}},\ and\ \citenamefont {{Terenziani}}}]{2009A&A...501...61E}%
  \BibitemOpen
  \bibfield  {author} {\bibinfo {author} {\bibfnamefont {S.}~\bibnamefont {{Ettori}}}, \bibinfo {author} {\bibfnamefont {A.}~\bibnamefont {{Morandi}}}, \bibinfo {author} {\bibfnamefont {P.}~\bibnamefont {{Tozzi}}}, \bibinfo {author} {\bibfnamefont {I.}~\bibnamefont {{Balestra}}}, \bibinfo {author} {\bibfnamefont {S.}~\bibnamefont {{Borgani}}}, \bibinfo {author} {\bibfnamefont {P.}~\bibnamefont {{Rosati}}}, \bibinfo {author} {\bibfnamefont {L.}~\bibnamefont {{Lovisari}}}, \ and\ \bibinfo {author} {\bibfnamefont {F.}~\bibnamefont {{Terenziani}}},\ }\href {\doibase 10.1051/0004-6361/200810878} {\bibfield  {journal} {\bibinfo  {journal} {aap}\ }\textbf {\bibinfo {volume} {501}},\ \bibinfo {pages} {61} (\bibinfo {year} {2009})},\ \Eprint {http://arxiv.org/abs/0904.2740} {arXiv:0904.2740 [astro-ph.CO]} \BibitemShut {NoStop}%
\bibitem [{\citenamefont {{Allen}}\ \emph {et~al.}(2011{\natexlab{b}})\citenamefont {{Allen}}, \citenamefont {{Evrard}},\ and\ \citenamefont {{Mantz}}}]{2011ARA&A..49..409A}%
  \BibitemOpen
  \bibfield  {author} {\bibinfo {author} {\bibfnamefont {S.~W.}\ \bibnamefont {{Allen}}}, \bibinfo {author} {\bibfnamefont {A.~E.}\ \bibnamefont {{Evrard}}}, \ and\ \bibinfo {author} {\bibfnamefont {A.~B.}\ \bibnamefont {{Mantz}}},\ }\href {\doibase 10.1146/annurev-astro-081710-102514} {\bibfield  {journal} {\bibinfo  {journal} {araa}\ }\textbf {\bibinfo {volume} {49}},\ \bibinfo {pages} {409} (\bibinfo {year} {2011}{\natexlab{b}})},\ \Eprint {http://arxiv.org/abs/1103.4829} {arXiv:1103.4829 [astro-ph.CO]} \BibitemShut {NoStop}%
\bibitem [{\citenamefont {Mantz}\ \emph {et~al.}(2014)\citenamefont {Mantz}, \citenamefont {Allen}, \citenamefont {Morris}, \citenamefont {Rapetti}, \citenamefont {Applegate}, \citenamefont {Kelly}, \citenamefont {von~der Linden},\ and\ \citenamefont {Schmidt}}]{Mantz:2014xba}%
  \BibitemOpen
  \bibfield  {author} {\bibinfo {author} {\bibfnamefont {A.~B.}\ \bibnamefont {Mantz}}, \bibinfo {author} {\bibfnamefont {S.~W.}\ \bibnamefont {Allen}}, \bibinfo {author} {\bibfnamefont {R.~G.}\ \bibnamefont {Morris}}, \bibinfo {author} {\bibfnamefont {D.~A.}\ \bibnamefont {Rapetti}}, \bibinfo {author} {\bibfnamefont {D.~E.}\ \bibnamefont {Applegate}}, \bibinfo {author} {\bibfnamefont {P.~L.}\ \bibnamefont {Kelly}}, \bibinfo {author} {\bibfnamefont {A.}~\bibnamefont {von~der Linden}}, \ and\ \bibinfo {author} {\bibfnamefont {R.~W.}\ \bibnamefont {Schmidt}},\ }\href {\doibase 10.1093/mnras/stu368} {\bibfield  {journal} {\bibinfo  {journal} {Mon. Not. Roy. Astron. Soc.}\ }\textbf {\bibinfo {volume} {440}},\ \bibinfo {pages} {2077} (\bibinfo {year} {2014})},\ \Eprint {http://arxiv.org/abs/1402.6212} {arXiv:1402.6212 [astro-ph.CO]} \BibitemShut {NoStop}%
\bibitem [{\citenamefont {{Holanda}}\ \emph {et~al.}(2019)\citenamefont {{Holanda}}, \citenamefont {{Gon{\c{c}}alves}}, \citenamefont {{Gonzalez}},\ and\ \citenamefont {{Alcaniz}}}]{2019JCAP...11..032H}%
  \BibitemOpen
  \bibfield  {author} {\bibinfo {author} {\bibfnamefont {R.~F.~L.}\ \bibnamefont {{Holanda}}}, \bibinfo {author} {\bibfnamefont {R.~S.}\ \bibnamefont {{Gon{\c{c}}alves}}}, \bibinfo {author} {\bibfnamefont {J.~E.}\ \bibnamefont {{Gonzalez}}}, \ and\ \bibinfo {author} {\bibfnamefont {J.~S.}\ \bibnamefont {{Alcaniz}}},\ }\href {\doibase 10.1088/1475-7516/2019/11/032} {\bibfield  {journal} {\bibinfo  {journal} {jcap}\ }\textbf {\bibinfo {volume} {2019}},\ \bibinfo {eid} {032} (\bibinfo {year} {2019})},\ \Eprint {http://arxiv.org/abs/1905.09689} {arXiv:1905.09689 [astro-ph.CO]} \BibitemShut {NoStop}%
\bibitem [{\citenamefont {Allen}\ \emph {et~al.}()\citenamefont {Allen}, \citenamefont {Rapetti}, \citenamefont {Schmidt}, \citenamefont {Ebeling}, \citenamefont {Morris},\ and\ \citenamefont {Fabian}}]{allen2008improved}%
  \BibitemOpen
  \bibfield  {author} {\bibinfo {author} {\bibfnamefont {S.}~\bibnamefont {Allen}}, \bibinfo {author} {\bibfnamefont {D.}~\bibnamefont {Rapetti}}, \bibinfo {author} {\bibfnamefont {R.}~\bibnamefont {Schmidt}}, \bibinfo {author} {\bibfnamefont {H.}~\bibnamefont {Ebeling}}, \bibinfo {author} {\bibfnamefont {R.}~\bibnamefont {Morris}}, \ and\ \bibinfo {author} {\bibfnamefont {A.}~\bibnamefont {Fabian}},\ }\href@noop {} {\ }\bibinfo {note} {\textit{Mon. Not. R. Astron. Soc.} \textbf{383}, 879-896 (2008)}\BibitemShut {NoStop}%
\bibitem [{\citenamefont {Planelles}\ \emph {et~al.}(2013)\citenamefont {Planelles}, \citenamefont {Borgani}, \citenamefont {Dolag}, \citenamefont {Ettori}, \citenamefont {Fabjan}, \citenamefont {Murante},\ and\ \citenamefont {Tornatore}}]{10.1093/mnras/stt265}%
  \BibitemOpen
  \bibfield  {author} {\bibinfo {author} {\bibfnamefont {S.}~\bibnamefont {Planelles}}, \bibinfo {author} {\bibfnamefont {S.}~\bibnamefont {Borgani}}, \bibinfo {author} {\bibfnamefont {K.}~\bibnamefont {Dolag}}, \bibinfo {author} {\bibfnamefont {S.}~\bibnamefont {Ettori}}, \bibinfo {author} {\bibfnamefont {D.}~\bibnamefont {Fabjan}}, \bibinfo {author} {\bibfnamefont {G.}~\bibnamefont {Murante}}, \ and\ \bibinfo {author} {\bibfnamefont {L.}~\bibnamefont {Tornatore}},\ }\href {\doibase 10.1093/mnras/stt265} {\bibfield  {journal} {\bibinfo  {journal} {Monthly Notices of the Royal Astronomical Society}\ }\textbf {\bibinfo {volume} {431}},\ \bibinfo {pages} {1487} (\bibinfo {year} {2013})},\ \Eprint {http://arxiv.org/abs/https://academic.oup.com/mnras/article-pdf/431/2/1487/4265857/stt265.pdf} {https://academic.oup.com/mnras/article-pdf/431/2/1487/4265857/stt265.pdf} \BibitemShut {NoStop}%
\bibitem [{\citenamefont {Applegate}\ \emph {et~al.}(2014)\citenamefont {Applegate}, \citenamefont {von~der Linden}, \citenamefont {Kelly}, \citenamefont {Allen}, \citenamefont {Allen}, \citenamefont {Burchat}, \citenamefont {Burke}, \citenamefont {Ebeling}, \citenamefont {Mantz},\ and\ \citenamefont {Morris}}]{10.1093/mnras/stt2129}%
  \BibitemOpen
  \bibfield  {author} {\bibinfo {author} {\bibfnamefont {D.~E.}\ \bibnamefont {Applegate}}, \bibinfo {author} {\bibfnamefont {A.}~\bibnamefont {von~der Linden}}, \bibinfo {author} {\bibfnamefont {P.~L.}\ \bibnamefont {Kelly}}, \bibinfo {author} {\bibfnamefont {M.~T.}\ \bibnamefont {Allen}}, \bibinfo {author} {\bibfnamefont {S.~W.}\ \bibnamefont {Allen}}, \bibinfo {author} {\bibfnamefont {P.~R.}\ \bibnamefont {Burchat}}, \bibinfo {author} {\bibfnamefont {D.~L.}\ \bibnamefont {Burke}}, \bibinfo {author} {\bibfnamefont {H.}~\bibnamefont {Ebeling}}, \bibinfo {author} {\bibfnamefont {A.}~\bibnamefont {Mantz}}, \ and\ \bibinfo {author} {\bibfnamefont {R.~G.}\ \bibnamefont {Morris}},\ }\href {\doibase 10.1093/mnras/stt2129} {\bibfield  {journal} {\bibinfo  {journal} {Monthly Notices of the Royal Astronomical Society}\ }\textbf {\bibinfo {volume} {439}},\ \bibinfo {pages} {48} (\bibinfo {year} {2014})},\ \Eprint {http://arxiv.org/abs/https://academic.oup.com/mnras/article-pdf/439/1/48/5560953/stt2129.pdf}
  {https://academic.oup.com/mnras/article-pdf/439/1/48/5560953/stt2129.pdf} \BibitemShut {NoStop}%
\bibitem [{\citenamefont {{Applegate}}\ \emph {et~al.}(2014)\citenamefont {{Applegate}}, \citenamefont {{von der Linden}}, \citenamefont {{Kelly}}, \citenamefont {{Allen}}, \citenamefont {{Allen}}, \citenamefont {{Burchat}}, \citenamefont {{Burke}}, \citenamefont {{Ebeling}}, \citenamefont {{Mantz}},\ and\ \citenamefont {{Morris}}}]{2014MNRAS.439...48A}%
  \BibitemOpen
  \bibfield  {author} {\bibinfo {author} {\bibfnamefont {D.~E.}\ \bibnamefont {{Applegate}}}, \bibinfo {author} {\bibfnamefont {A.}~\bibnamefont {{von der Linden}}}, \bibinfo {author} {\bibfnamefont {P.~L.}\ \bibnamefont {{Kelly}}}, \bibinfo {author} {\bibfnamefont {M.~T.}\ \bibnamefont {{Allen}}}, \bibinfo {author} {\bibfnamefont {S.~W.}\ \bibnamefont {{Allen}}}, \bibinfo {author} {\bibfnamefont {P.~R.}\ \bibnamefont {{Burchat}}}, \bibinfo {author} {\bibfnamefont {D.~L.}\ \bibnamefont {{Burke}}}, \bibinfo {author} {\bibfnamefont {H.}~\bibnamefont {{Ebeling}}}, \bibinfo {author} {\bibfnamefont {A.}~\bibnamefont {{Mantz}}}, \ and\ \bibinfo {author} {\bibfnamefont {R.~G.}\ \bibnamefont {{Morris}}},\ }\href {\doibase 10.1093/mnras/stt2129} {\bibfield  {journal} {\bibinfo  {journal} {mnras}\ }\textbf {\bibinfo {volume} {439}},\ \bibinfo {pages} {48} (\bibinfo {year} {2014})},\ \Eprint {http://arxiv.org/abs/1208.0605} {arXiv:1208.0605 [astro-ph.CO]} \BibitemShut {NoStop}%
\bibitem [{\citenamefont {Henden}\ \emph {et~al.}(2020)\citenamefont {Henden}, \citenamefont {Puchwein},\ and\ \citenamefont {Sijacki}}]{10.1093/mnras/staa2235}%
  \BibitemOpen
  \bibfield  {author} {\bibinfo {author} {\bibfnamefont {N.~A.}\ \bibnamefont {Henden}}, \bibinfo {author} {\bibfnamefont {E.}~\bibnamefont {Puchwein}}, \ and\ \bibinfo {author} {\bibfnamefont {D.}~\bibnamefont {Sijacki}},\ }\href {\doibase 10.1093/mnras/staa2235} {\bibfield  {journal} {\bibinfo  {journal} {Monthly Notices of the Royal Astronomical Society}\ }\textbf {\bibinfo {volume} {498}},\ \bibinfo {pages} {2114} (\bibinfo {year} {2020})},\ \Eprint {http://arxiv.org/abs/https://academic.oup.com/mnras/article-pdf/498/2/2114/33776942/staa2235.pdf} {https://academic.oup.com/mnras/article-pdf/498/2/2114/33776942/staa2235.pdf} \BibitemShut {NoStop}%
\bibitem [{\citenamefont {Herbonnet}\ \emph {et~al.}(2020)\citenamefont {Herbonnet}, \citenamefont {Sifón}, \citenamefont {Hoekstra}, \citenamefont {Bahé}, \citenamefont {van der Burg}, \citenamefont {Melin}, \citenamefont {von der Linden}, \citenamefont {Sand}, \citenamefont {Kay},\ and\ \citenamefont {Barnes}}]{10.1093/mnras/staa2303}%
  \BibitemOpen
  \bibfield  {author} {\bibinfo {author} {\bibfnamefont {R.}~\bibnamefont {Herbonnet}}, \bibinfo {author} {\bibfnamefont {C.}~\bibnamefont {Sifón}}, \bibinfo {author} {\bibfnamefont {H.}~\bibnamefont {Hoekstra}}, \bibinfo {author} {\bibfnamefont {Y.}~\bibnamefont {Bahé}}, \bibinfo {author} {\bibfnamefont {R.~F.~J.}\ \bibnamefont {van der Burg}}, \bibinfo {author} {\bibfnamefont {J.-B.}\ \bibnamefont {Melin}}, \bibinfo {author} {\bibfnamefont {A.}~\bibnamefont {von der Linden}}, \bibinfo {author} {\bibfnamefont {D.}~\bibnamefont {Sand}}, \bibinfo {author} {\bibfnamefont {S.}~\bibnamefont {Kay}}, \ and\ \bibinfo {author} {\bibfnamefont {D.}~\bibnamefont {Barnes}},\ }\href {\doibase 10.1093/mnras/staa2303} {\bibfield  {journal} {\bibinfo  {journal} {Monthly Notices of the Royal Astronomical Society}\ }\textbf {\bibinfo {volume} {497}},\ \bibinfo {pages} {4684} (\bibinfo {year} {2020})},\ \Eprint {http://arxiv.org/abs/https://academic.oup.com/mnras/article-pdf/497/4/4684/33680798/staa2303.pdf}
  {https://academic.oup.com/mnras/article-pdf/497/4/4684/33680798/staa2303.pdf} \BibitemShut {NoStop}%
\bibitem [{\citenamefont {Hoekstra}\ \emph {et~al.}(2015)\citenamefont {Hoekstra}, \citenamefont {Herbonnet}, \citenamefont {Muzzin}, \citenamefont {Babul}, \citenamefont {Mahdavi}, \citenamefont {Viola},\ and\ \citenamefont {Cacciato}}]{10.1093/mnras/stv275}%
  \BibitemOpen
  \bibfield  {author} {\bibinfo {author} {\bibfnamefont {H.}~\bibnamefont {Hoekstra}}, \bibinfo {author} {\bibfnamefont {R.}~\bibnamefont {Herbonnet}}, \bibinfo {author} {\bibfnamefont {A.}~\bibnamefont {Muzzin}}, \bibinfo {author} {\bibfnamefont {A.}~\bibnamefont {Babul}}, \bibinfo {author} {\bibfnamefont {A.}~\bibnamefont {Mahdavi}}, \bibinfo {author} {\bibfnamefont {M.}~\bibnamefont {Viola}}, \ and\ \bibinfo {author} {\bibfnamefont {M.}~\bibnamefont {Cacciato}},\ }\href {\doibase 10.1093/mnras/stv275} {\bibfield  {journal} {\bibinfo  {journal} {Monthly Notices of the Royal Astronomical Society}\ }\textbf {\bibinfo {volume} {449}},\ \bibinfo {pages} {685} (\bibinfo {year} {2015})},\ \Eprint {http://arxiv.org/abs/https://academic.oup.com/mnras/article-pdf/449/1/685/4136034/stv275.pdf} {https://academic.oup.com/mnras/article-pdf/449/1/685/4136034/stv275.pdf} \BibitemShut {NoStop}%
\bibitem [{\citenamefont {Angelinelli}\ \emph {et~al.}(2022)\citenamefont {Angelinelli}, \citenamefont {Ettori}, \citenamefont {Dolag}, \citenamefont {Vazza},\ and\ \citenamefont {Ragagnin}}]{Angelinelli:2022njb}%
  \BibitemOpen
  \bibfield  {author} {\bibinfo {author} {\bibfnamefont {M.}~\bibnamefont {Angelinelli}}, \bibinfo {author} {\bibfnamefont {S.}~\bibnamefont {Ettori}}, \bibinfo {author} {\bibfnamefont {K.}~\bibnamefont {Dolag}}, \bibinfo {author} {\bibfnamefont {F.}~\bibnamefont {Vazza}}, \ and\ \bibinfo {author} {\bibfnamefont {A.}~\bibnamefont {Ragagnin}},\ }\href {\doibase 10.1051/0004-6361/202244068} {\bibfield  {journal} {\bibinfo  {journal} {Astron. Astrophys.}\ }\textbf {\bibinfo {volume} {663}},\ \bibinfo {pages} {L6} (\bibinfo {year} {2022})},\ \Eprint {http://arxiv.org/abs/2206.08382} {arXiv:2206.08382 [astro-ph.GA]} \BibitemShut {NoStop}%
\bibitem [{\citenamefont {Riess}\ \emph {et~al.}(2016)\citenamefont {Riess} \emph {et~al.}}]{Riess:2016jrr}%
  \BibitemOpen
  \bibfield  {author} {\bibinfo {author} {\bibfnamefont {A.~G.}\ \bibnamefont {Riess}} \emph {et~al.},\ }\href {\doibase 10.3847/0004-637X/826/1/56} {\bibfield  {journal} {\bibinfo  {journal} {Astrophys. J.}\ }\textbf {\bibinfo {volume} {826}},\ \bibinfo {pages} {56} (\bibinfo {year} {2016})},\ \Eprint {http://arxiv.org/abs/1604.01424} {arXiv:1604.01424 [astro-ph.CO]} \BibitemShut {NoStop}%
\bibitem [{\citenamefont {Seikel}\ \emph {et~al.}(2012)\citenamefont {Seikel}, \citenamefont {Clarkson},\ and\ \citenamefont {Smith}}]{Seikel_2012}%
  \BibitemOpen
  \bibfield  {author} {\bibinfo {author} {\bibfnamefont {M.}~\bibnamefont {Seikel}}, \bibinfo {author} {\bibfnamefont {C.}~\bibnamefont {Clarkson}}, \ and\ \bibinfo {author} {\bibfnamefont {M.}~\bibnamefont {Smith}},\ }\href {\doibase 10.1088/1475-7516/2012/06/036} {\bibfield  {journal} {\bibinfo  {journal} {Journal of Cosmology and Astroparticle Physics}\ }\textbf {\bibinfo {volume} {2012}},\ \bibinfo {pages} {036–036} (\bibinfo {year} {2012})}\BibitemShut {NoStop}%
\bibitem [{\citenamefont {{Foreman-Mackey}}\ \emph {et~al.}(2013)\citenamefont {{Foreman-Mackey}}, \citenamefont {{Hogg}}, \citenamefont {{Lang}},\ and\ \citenamefont {{Goodman}}}]{2013PASP125306F}%
  \BibitemOpen
  \bibfield  {author} {\bibinfo {author} {\bibfnamefont {D.}~\bibnamefont {{Foreman-Mackey}}}, \bibinfo {author} {\bibfnamefont {D.~W.}\ \bibnamefont {{Hogg}}}, \bibinfo {author} {\bibfnamefont {D.}~\bibnamefont {{Lang}}}, \ and\ \bibinfo {author} {\bibfnamefont {J.}~\bibnamefont {{Goodman}}},\ }\href {\doibase 10.1086/670067} {\bibfield  {journal} {\bibinfo  {journal} {pasp}\ }\textbf {\bibinfo {volume} {125}},\ \bibinfo {pages} {306} (\bibinfo {year} {2013})},\ \Eprint {http://arxiv.org/abs/1202.3665} {arXiv:1202.3665 [astro-ph.IM]} \BibitemShut {NoStop}%
\bibitem [{\citenamefont {Lewis}(2019)}]{Lewis:2019xzd}%
  \BibitemOpen
  \bibfield  {author} {\bibinfo {author} {\bibfnamefont {A.}~\bibnamefont {Lewis}},\ }\href@noop {} {\  (\bibinfo {year} {2019})},\ \Eprint {http://arxiv.org/abs/1910.13970} {arXiv:1910.13970 [astro-ph.IM]} \BibitemShut {NoStop}%
\bibitem [{\citenamefont {Schmelling}(2000)}]{Schmelling:2000td}%
  \BibitemOpen
  \bibfield  {author} {\bibinfo {author} {\bibfnamefont {M.}~\bibnamefont {Schmelling}},\ }\href@noop {} {\  (\bibinfo {year} {2000})},\ \Eprint {http://arxiv.org/abs/hep-ex/0006004} {arXiv:hep-ex/0006004} \BibitemShut {NoStop}%
\end{thebibliography}%

\end{document}